%% file: methodology.tex
\documentclass[useAMS,usenatbib]{mn2e}
\usepackage{epsfig}

\usepackage{lipsum}
\usepackage{lscape}
\usepackage{longtable,amsmath}
\usepackage{booktabs}

\title[Cluster properties]{Properties of Star Clusters - I: Automatic
Distance and Extinction Estimates}

\author[Buckner \& Froebrich]{Anne~S.M. Buckner$^{1}$\thanks{E-mail:
asmb2@kent.ac.uk}, Dirk Froebrich$^{1}$\thanks{E-mail: df@star.kent.ac.uk}\\
$^{1}$ Centre for Astrophysics and Planetary Science, University of Kent,
Canterbury, CT2 7NH, United Kingdom }

\begin{document}

\date{Accepted. Received.}

\pagerange{\pageref{firstpage}--\pageref{lastpage}} \pubyear{2012}

\maketitle

\label{firstpage}

\begin{abstract}

Determining star cluster distances is essential to analyse their properties and
distribution in the Galaxy. In particular it is desirable to have a reliable,
purely photometric distance estimation method for large samples of newly
discovered cluster candidates e.g. from 2MASS, UKIDSS-GPS and VISTA-VVV. Here,
we establish an automatic method to estimate distances and reddening from NIR
photometry alone, without the use of isochrone fitting. We employ a
decontamination procedure of JHK photometry to determine the density of stars
foreground to clusters and a galactic model to estimate distances. We then
calibrate the method using clusters with known properties. This allows us to
establish distance estimates with better than 40\,\% accuracy.

We apply our method to determine the extinction and distance values to 378
known open clusters and 397 cluster candidates from the list of Froebrich,
Scholz and Raftery (2003). We find that the sample is biased towards clusters
of a distance of approximately 3\,kpc, with typical distances between 2 and
6\,kpc. Using the cluster distances and extinction values, we investigate how
the average extinction per kiloparsec distance changes as a function of
Galactic longitude. We find a systematic dependence that can be approximated by
$A_H(l)$\,[mag/kpc]\,=\,0.10\,+\,0.001$\times \left| l - 180^\circ \right|
/^\circ$ for regions more than 60$^\circ$ from the Galactic Centre.

\end{abstract}


\section{Introduction}
\label{intro}
 
Star clusters are the building blocks of the Galaxy and act as tracers of
stellar and Galactic evolution. With the large fraction of stars in the Milky
Way formed in star clusters (e.g. \citet{Lada2003}), it is important to
determine fundamental properties of clusters such as age, reddening and mass.
Determination of distance is essential to analyse these properties and the
distribution of clusters throughout the Galaxy. 

When a large sample of clusters is available, objects of interest such as
massive clusters and old clusters near the Galactic Centre (GC, see e.g.
\citet{2007A&A...473..445B}), become available to study. Large cluster
candidate samples, such as the list based on data from the 2\,Micron All Sky
Survey (2MASS) by \citet{2007MNRAS.374..399F} (FSR hereafter) and others (e.g.
\citet{2011A&A...532A.131B}, \citet{2012A&A...545A..54C}) have become readily
available in recent years. Further samples are forthcoming from large scale
Near Infrared (NIR) surveys such as the UK Infrared Deep Sky Survey (UKIDSS)
Galactic Plane Survey (GPS, \citet{Lucas2008}) and the VISTA-VVV survey
\citep{Minniti2010}. Here we aim to establish an automatic method to estimate
distances and reddening for such large cluster candidate samples from NIR
photometry alone without the use of, or to be used as starting point for,
isochrone fitting.

In a forthcoming paper (Paper\,II) we will extend our work and determine the ages
of all clusters. We will also improve the accuracy of the distances and
extinction estimates from this paper by means of isochrone fitting. That will
provide us with the ability to characterise and analyse the general properties of
the entire FSR sample and to investigate the distribution of ages, reddening and
distances as well as their spatial distribution, observational biases and
evolutionary trends. It will also allow us to extract the best massive cluster
candidates amongst the FSR sample.

This paper is structured as follows. In Section\,\ref{method} we discuss our
selections of 2MASS$/$WISE data and describe in detail our foreground star
counting technique as a distance estimator. We then introduce and detail our
distance calibration and optimisation method in Section\,\ref{distcal}. In
Section\,\ref{results} we discuss the results of our calibration and optimisation
method applied to the FSR list of cluster candidates and the extinction $A_H$ per
kpc measurement as a function of Galactic longitude.

\section{Analysis Methods}
\label{method}

In the following section we will detail our homogeneous and automated approach to
estimate and calibrate distances and extinction values for all FSR clusters,
based on photometric archival data. Our suggested method can be applied to any
large sample of star clusters and cluster candidates containing a sufficient
number of objects with known distances (which will be used as calibrators).

\subsection{2MASS/WISE Data and Cluster Radii}\label{find_radii}
 
Our distance and extinction estimates will rely on identifying the colour of the
most likely cluster members. For this purpose we will determine a
membership-likelihood or photometric cluster membership probability for every
star in each cluster (see Sect.\,\ref{cluster_prob}). As a first step we hence
need to specify the area around the cluster which contains most cluster members,
as well as an area near the cluster which will be used as control field.

For this purpose we use the coordinates of all FSR clusters as published in
\citet{2007MNRAS.374..399F}. Note that this paper only contains the coordinates
of the 1021 newly discovered cluster candidates. The positions of the remaining
681 previously known open clusters and 86 globular clusters are not published. 
For each FSR cluster we extract near infrared JHK photometry from the 2MASS
\citep{2006AJ....131.1163S} point source catalogue. All sources within a
circular area of $0.5^\circ$ radius around the cluster coordinates are
selected, as long as the photometric quality flag (Qflg) was better than "CCC".
Typically, between 50\,\% and 70\,\% of all stars in the catalogue have a Qflg
better than "CCC" (C-sample, hereafter). Furthermore, about 35\,\% to 45\,\% of
all stars have a Qflg of "AAA" (A-sample, hereafter), i.e. are of the highest
photometric quality. Thus, the C-sample contains on average about 1.5 times as
many stars as the A-sample.

For each cluster we use the C-sample to fit a radial star density profile of the
form:

\begin{equation}\label{eq_density}
\rho(r)=\rho_{bg}+\rho_{cen} \left[ 1 + \left(\frac{r}{r_{cor}}\right)^2
\right]^{-1}
\end{equation}

Where $r_{cor}$ is the cluster core radius, $\rho(r)$ the star density as a
function of distance $r$ from the cluster centre, $\rho_{bg}$ the (assumed
constant) background star density and $\rho_{cen}$ the central star density
above the background of the cluster.

We define as cluster area ($A_{cl}$) everything in a circular area around the
cluster centre within $F$ times the cluster core radius. In our subsequent
analysis we will vary the value of $F$ between one and three. The control area
($A_{con}$) for each cluster will be a ring around the cluster centre with an
inner radius of five times $r_{cor}$ and an outer radius of $0.5^\circ$.

Note that the star density profile used has no tidal radius and thus in
principle results in an infinite number of cluster stars. However, if we assume
that outside five core radii (the control field) there are no member stars, then
the region within three core radii contains about 70\,\% of all cluster members.
If all cluster stars are contained within three core radii, then 70\,\% of the
cluster members are found closer than two core radii from the centre.

In additional to the 2MASS Near Infrared photometry, we also utilise mid infrared
data from the WISE satellite in order to estimate the colour excess of cluster
stars and hence the extinction (see Sect.\,\ref{ext_cal}). We obtain all sources
from the WISE all-sky catalogue \citep{2010AJ....140.1868W} within three times
the cluster core radius. The WISE all-sky catalogue is cross-matched to 2MASS,
hence we can easily identify the stars in both, the A- and C-sample, that have a
WISE counterpart.

\subsection{Photometric Cluster Membership Probabilities}\label{cluster_prob}
 
In order to estimate the typical colour of cluster members for our analysis, as
well as to identify potential foreground and background stars to determine the
distance, we require some measure of membership-likelihood for every star in a
cluster. We base our calculation on the well established NIR
colour-colour-magnitude (CCM) method outlined in e.g.
\citet{2007MNRAS.377.1301B} and references therein. Based on earlier
simulations, e.g. in \citet{2004A&A...415..571B}, \citet{2007MNRAS.377.1301B}
note that using the J-band magnitude, as well as the J-H and J-K colours from
2MASS photometry, provides the maximum variance among cluster colour-magnitude
sequences for open clusters of different ages. Since our sample will most likely
contain clusters of all  ages, we thus use the same colours for our analysis.

All confirmed clusters and cluster candidates in the FSR list are selected as
spatial overdensities. Our photometric cluster membership probabilities are
based on 'local' overdensities in the above mentioned CCM space. Thus, we need
to establish where these overdensities are in CCM space and if these
overdensities are in agreement with the expectation for a real star cluster.
\citet{2007MNRAS.377.1301B} use a small cuboid in CCM space to determine the
overdensities and the related photometric membership probabilities, where the
dimension of the J-magnitude side length of the cuboid is larger than the side
length of the colours. Instead of a small cuboid, \citet{2010MNRAS.409.1281F}
use a small prolate ellipsoid in the specified CCM space, where the dimensions
along the J-band magnitude are larger than along the colours. Note that the
actual shape used to determine the local overdensity in CCM space (cuboid,
ellipsoid, or else) is unimportant for the identification of the most likely
cluster members.

Thus, for each cluster we determine the photometric cluster membership
probability for every star using the method described below. Following
\citet{2010MNRAS.409.1281F} we determine the CCM distance, $r_{ccm}$, between
the star, $i$, and every other star $j \ne i$ in the cluster area using:

\begin{equation}\label{eq_rccm}
r_{ccm}=\sqrt{\frac{1}{2} \left( J_i - J_j \right)^2 + \left( JK_i - JK_j
\right)^2 + \left( JH_i - JH_j \right)^2}
\end{equation}

Where $JK = J-K$ and $JH = J-H$ are the Near Infrared colours from 2MASS for all
stars. We denote with $r^N_{ccm}$ the CCM distance of star $i$ to the $N^{th}$
nearest neighbour in CCM space. \citet{2010MNRAS.409.1281F}  used $N = 10$ to
investigate old star clusters amongst the FSR sample. However, it is not
formally established which value for $N$ gives the best results for our purpose.
We hence vary $N$ from 10 to 30 and investigate the influence of the value on
the distance calibration in Sect.\,\ref{model_parameters}. The value of $N$
essentially defines, in conjunction with the total number of stars in the
cluster area, the resolution at which we can separate potential cluster members
from field stars in CCM space. Increasing $N$ will degrade the resolution and
thus e.g. widen any potential cluster Main Sequence. Using smaller values for
$N$ will increase the resolution, but will at the same time decrease the signal
to noise ratio of the determined photometric cluster membership probabilities. 
Our range of values for $N$ is hence a compromise between resolution and signal
to noise ratio. Note, that the typical number of stars in the cluster area for
our objects is between 100 and 300 for the A-sample. Thus, our resolution in CCM
space varies by less than a factor of $1.5$ between clusters.

We then determine the CCM distance of all stars in the cluster control field to
star $i$ in the same way as for the cluster area. $N^{con}_{ccm}$ is then the
number of stars closer than $r^N_{ccm}$ to star $i$ in the control field. If
both $N$ and $N^{con}_{ccm}$ are normalised by their respective area on the sky,
one can calculate a membership-likelihood index of star $i$, $P^i_{cl}$ as:

\begin{equation}\label{eq_pcl}
P^i_{cl}=1.0-\frac{N^{con}_{ccm}}{N}\frac{A_{cl}}{A_{con}}.
\end{equation}

Due to statistical fluctuations in the number of field stars in the control and
cluster area, the above equation can in principle lead to negative values. We
thus set any negative $P^i_{cl}$ value to zero and note that $P^i_{cl}$ is in
principle not a real membership probability. However, all we require is a list
of the most likely cluster members, which are reliably identified by this
method. Note that after our calibration (see Sect.\,\ref{correct_overcrowding})
the sum of all $P^i_{cl}$ values equals the total excess of stars in the cluster
field compared to the control field. Furthermore, the sum all $1 - P^i_{cl}$
values equals the number of field stars. Thus, this membership-likelihood index
can be treated as a probability. Hence we will refer to $P^i_{cl}$ as the
photometric cluster membership probability of star $i$ hereafter.

In principle one could combine the above described colour based membership
probabilities with spatial information. In other words one could determine a
probability $P^i_{pos}$ based on the distance of the star to the cluster centre
as well as the background and cluster star density determined via
Eq.\,\ref{eq_density}. 

\begin{equation}\label{eq_pos}
P^i_{pos}=\frac{\rho_{cen}}{\rho_{bg}} \left[ 1 +
\left(\frac{r}{r_{cor}}\right)^2 \right]^{-1}
\end{equation}

Following \citet{2010MNRAS.409.1281F}, we refrain from applying this method for
several reasons: 

i) In dense clusters the probabilities are unreliable due to crowding near the
centre. 

ii) For clusters projected onto a high background density of stars the
$P^i_{pos}$ values tend to be very low. This makes the cluster not stand out
from the background stars in many cases, even if cluster stars have very
different colours compared to the field population.

iii) Our sample will contain a number of young clusters which do not appear
circular in projection. Thus, the position dependent membership probabilities
cannot be determined by Eq.\,\ref{eq_pos}. 

We hence only use the  $P^i_{cl}$ values to identify the most likely cluster
members.

Note that the individual photometric cluster membership probabilities are a
function of the 'free' parameters in our approach. Hence, $P^i_{cl}$ will
depend on the sample of stars (A- or C-sample) used, the cluster area $A_{cl}$
(between one and three cluster core radii), and the nearest neighbour $10 \le N
\le 30$ chosen in the CCM distance calculation. We will discuss this in detail
in Sect.\,\ref{model_parameters}. 

\begin{figure*}
\includegraphics[width=8.6cm,angle=0]{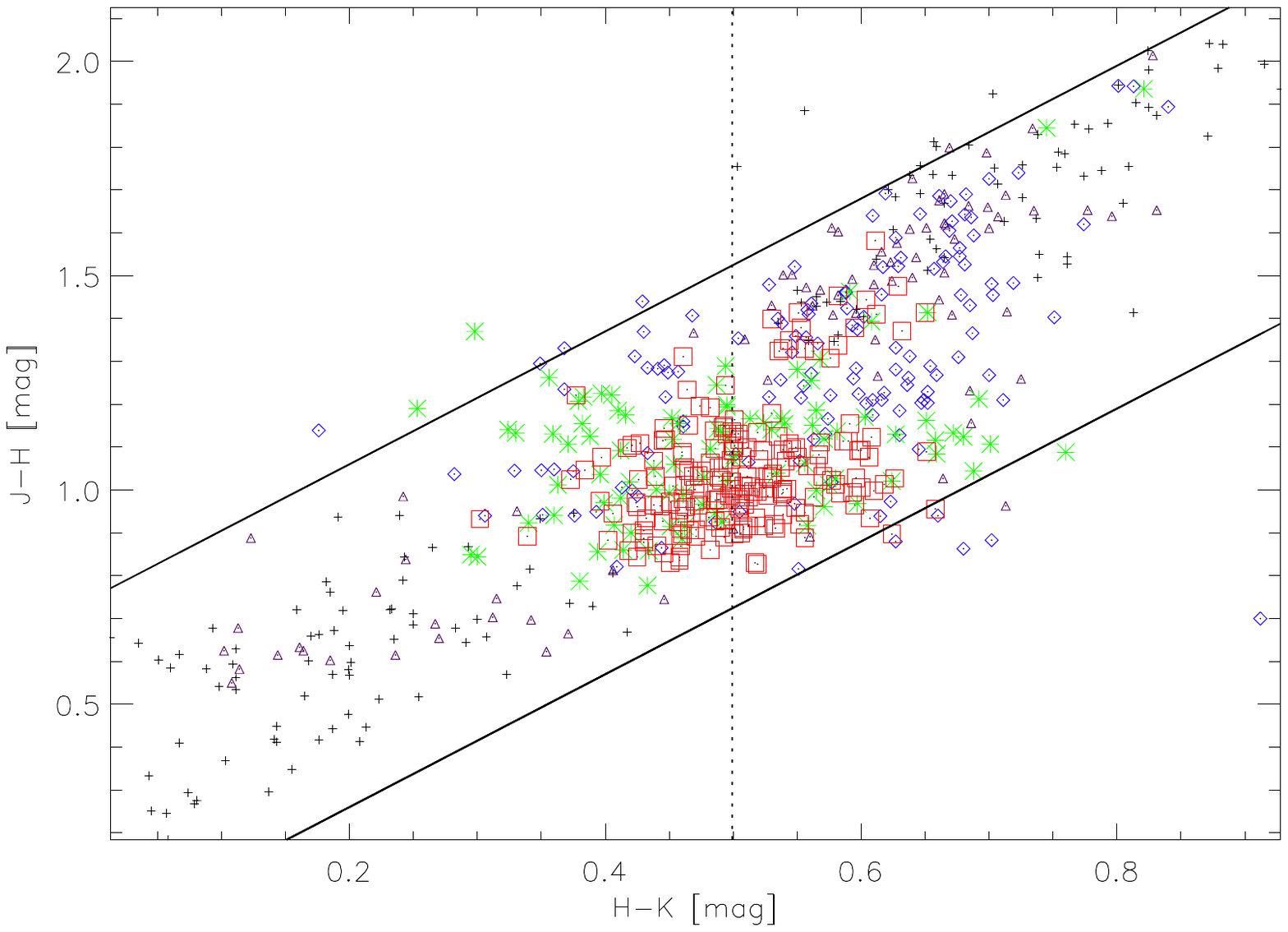}\hfill
\includegraphics[width=8.6cm,angle=0]{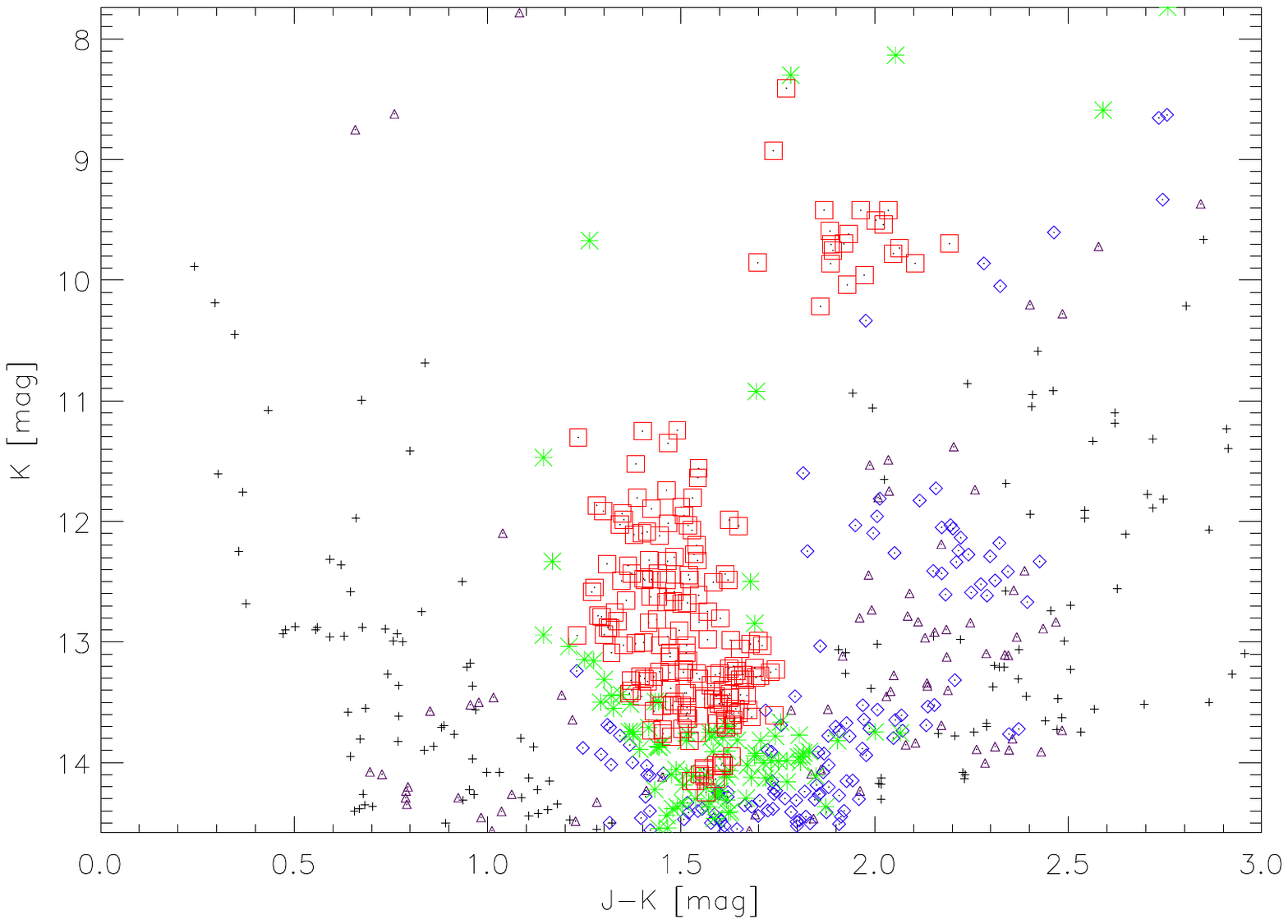}\\
\includegraphics[width=8.6cm,height=8.6cm,angle=0]{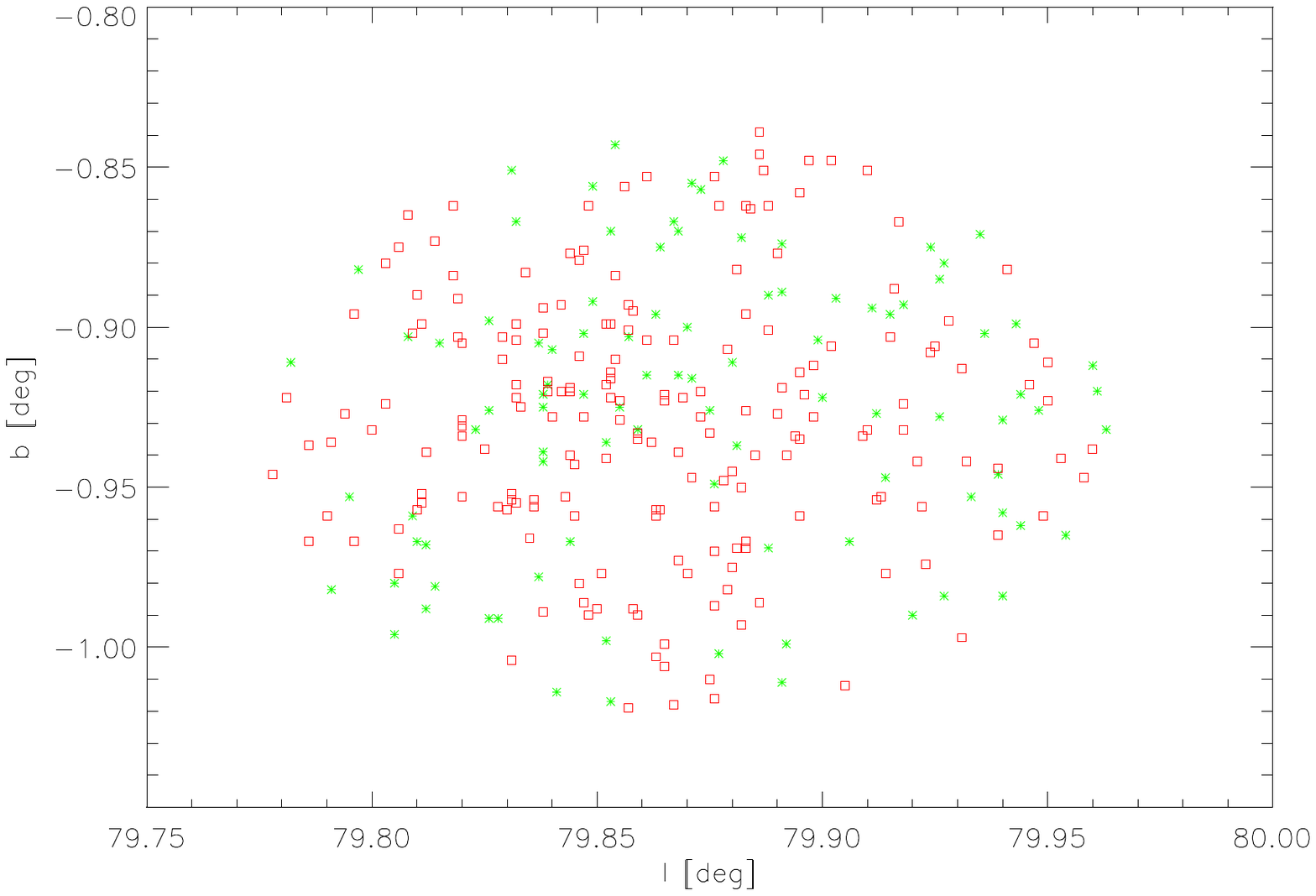}\hfill
\includegraphics[width=8.6cm,height=8.6cm,angle=0]{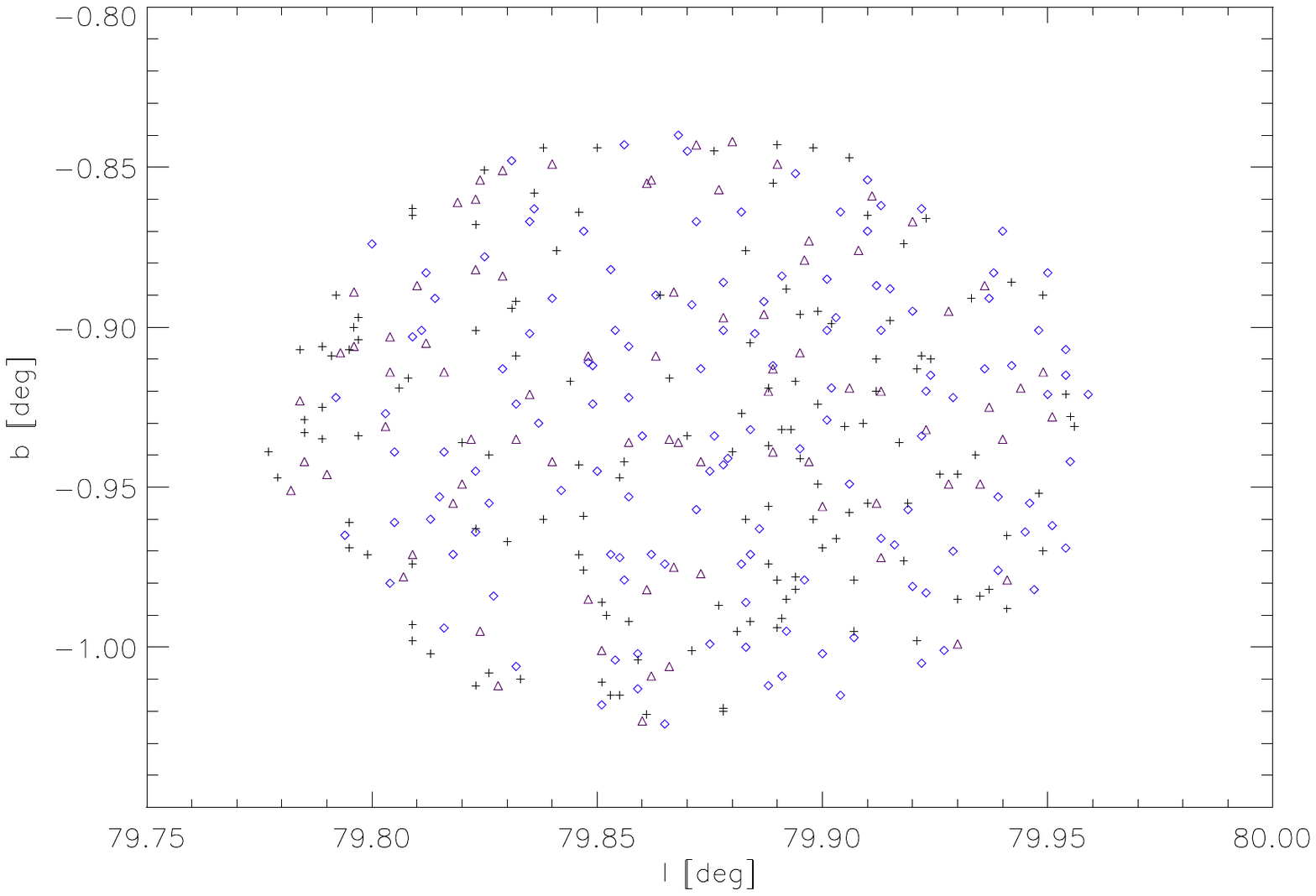}\\

\caption{\label{CCD_fsr0233} Diagrams of all A-sample stars within two cluster
core radii for FSR\,0233. The various symbols represent the cluster membership
probabilities determined for $N = 15$: $P^i_{cl} > 80\%$ red squares; $60 \leq
P^i_{cl} < 80\% $ green stars; $40 \leq P^i_{cl} < 60\% $ blue diamonds; $20
\leq P^i_{cl} < 40\% $ purple triangles; $P^i_{cl} < 20\% $ black plus signs.
The top row contains the colour-colour (left) and colour-magnitude diagram
(right). The vertical solid line in the left panel represents the $HK_{med}$
value (0.50\,mag) determined for the cluster (see text for details). In the
bottom row we plot the positions of the stars with a photometric membership
probability of more (left) and less (right) than 60\,\%. }

\end{figure*}

As an example of how our method performs, we show in Fig.\,\ref{CCD_fsr0233} 
several diagrams for the cluster FSR\,0233. Different coloured symbols in the
graphs indicate the various photometric cluster membership probabilities for the
stars determined for the A-sample within two cluster core radii and for $N =
15$. The top row shows the Colour-Colour Diagram (CCD, left) and
Colour-Magnitude Diagram (CMD, right) of the cluster region. In the CMD one can
clearly identify that the high probability cluster members form the top of a
Main Sequence and a clump of red giants, suggesting an older cluster. Indeed,
this is the known open cluster LK\,10, which is about 1\,Gyr old and relatively
massive \citep{2009MNRAS.392..483B}. In the bottom panels we show the spatial
distribution of the cluster stars. In the left panel we only include stars with
a membership probability above 60\,\%, while in the right panel we only plot the
remaining low probability members. One can clearly see that the low probability
members, which are the most likely field stars, are distributed homogeneously in
the field. In the left panel one can identify an increase in the density of
stars, slightly off-centre, that indicates the cluster centre.

\subsection{Identification of Foreground Stars}\label{id_foreground}

For our distance calculation it is not sufficient to solely identify the most likely
cluster members. We further need to establish the most likely foreground stars to
the cluster. 

We utilise the median colour of the stars with the highest cluster membership
probability to separate foreground and background objects. The notion is that
stars bluer than the median colour of the cluster members are likely to be
foreground and objects redder than this are most likely in the background. 

The combination of 2MASS and WISE allows us, in principle, to use a variety of
colours for this selection. However, ideally the colour with the smallest spread
amongst the spectral types and luminosity classes should be used, since any
reddening can then be attributed to interstellar extinction. The ideal choice
would be the difference between the 2MASS H-band and the 4.5\,$\mu$m WISE band
($H - [4.5]$), since its intrinsic value is almost independent of spectral type
and/or luminosity class \citep{2011ApJ...739...25M}. However, due to the low
spatial resolution of WISE, only a fraction of all 2MASS sources in each cluster
are actually detected unambiguously in $[4.5]$. We hence, use ($H-K$) for the
purpose of foreground star selection, but refer back to ($H - [4.5]$) in the
extinction determination (see Sect.\,\ref{ext_cal}).

For every cluster we arrange the individual photometric cluster membership
probabilities, $P^i_{cl}$, in descending order. The median ($H-K$) of the top
25\,\%\,--\,45\,\% of the highest probability members is calculated in 5\,\%
increments. We then define $HK_{med}$ as the median of these five ($H-K$)
values, to represent the most likely ($H-K$) colour of the cluster stars.  All
stars in the cluster area with an ($H-K$) value equal to, or greater than,
$HK_{med}$ are considered redder than the cluster and are hence potential
background stars. All stars with an ($H-K$) value less than $HK_{med}$ are
considered bluer than the cluster, and hence can be potential foreground
objects. Based on this, we determine the number density of foreground stars,
$\rho^{cl}_{fg}$, projected onto the cluster as the sum of all non-membership
probabilities of the potential 'blue' foreground stars:

\begin{equation}\label{eq_forg}
\rho^{cl}_{fg}=  \frac{1}{A_{cl}} \sum\limits_{i, blue}^{ }{ \left( 1.0 - P^i_{cl}
\right) }
\end{equation}

We note that for particular cases, where cluster members are intrinsically red,
such as embedded clusters, or clusters behind multiple layers of foreground
extinction, this method will not lead to the correct number of foreground stars.
The same applies if there are intrinsically hot/blue stars in the cluster area.
Furthermore, photometric scatter will influence the number of blue objects.
However, our calibration procedure (see Sect.\,\ref{distcal}) will statistically
adjust for this.

\subsection{Determination of Cluster Distances}\label{distance_det}

The above determined projected number density of foreground stars towards each
cluster can be used to estimate the distance, by comparing this number to
predictions from models for the distribution of stars within the Galaxy. Our
model of choice is the Besan\c{c}on Galaxy Model (BGM) by
\citet{2003A&A...409..523R} and we utilise its web interface\footnote{\tt
http://model.obs-besancon.fr/} to simulate the foreground population of stars
towards all our clusters. 

We use the default settings of the BGM. i.e. no clouds and 0.7\,mag of
optical extinction per kiloparsec distance. The photometric limits (completeness
limit and photometric uncertainties as a function of brightness) in the 2MASS
JHK filters are determined for each cluster and sample (A and C) separately in
the control field area. After initial tests with the above settings, we perform
simulations for each cluster position for an area large enough to contain a
total of about 5000 simulated stars. This ensures that the uncertainties of our
inferred distances are not dominated by the random nature (or small number
statistics) in the model output.

The list of stars returned by the BGM simulation for the field of the cluster is
sorted by distance. With the known area of the simulation, we are hence able to
determine up to which distance model stars need to be considered to result in a
model star density equal to our determined foreground density for the cluster.
Henceforth, we will refer to this distance to the cluster as the {\it Model
Distance}, $d_{mod}$. 

\begin{figure*}
\includegraphics[width=8.6cm,angle=0]{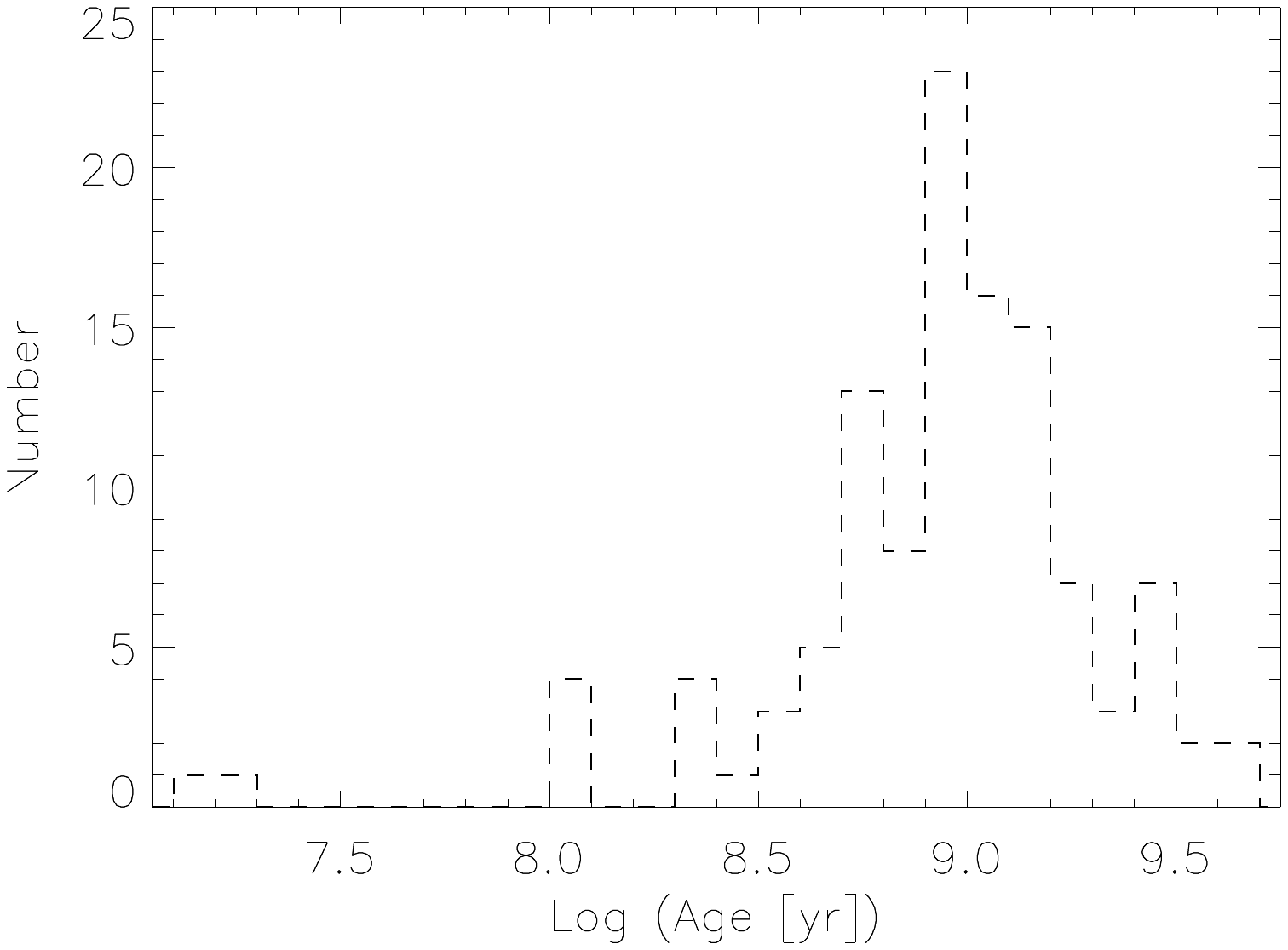} \hfill
\includegraphics[width=8.6cm,angle=0]{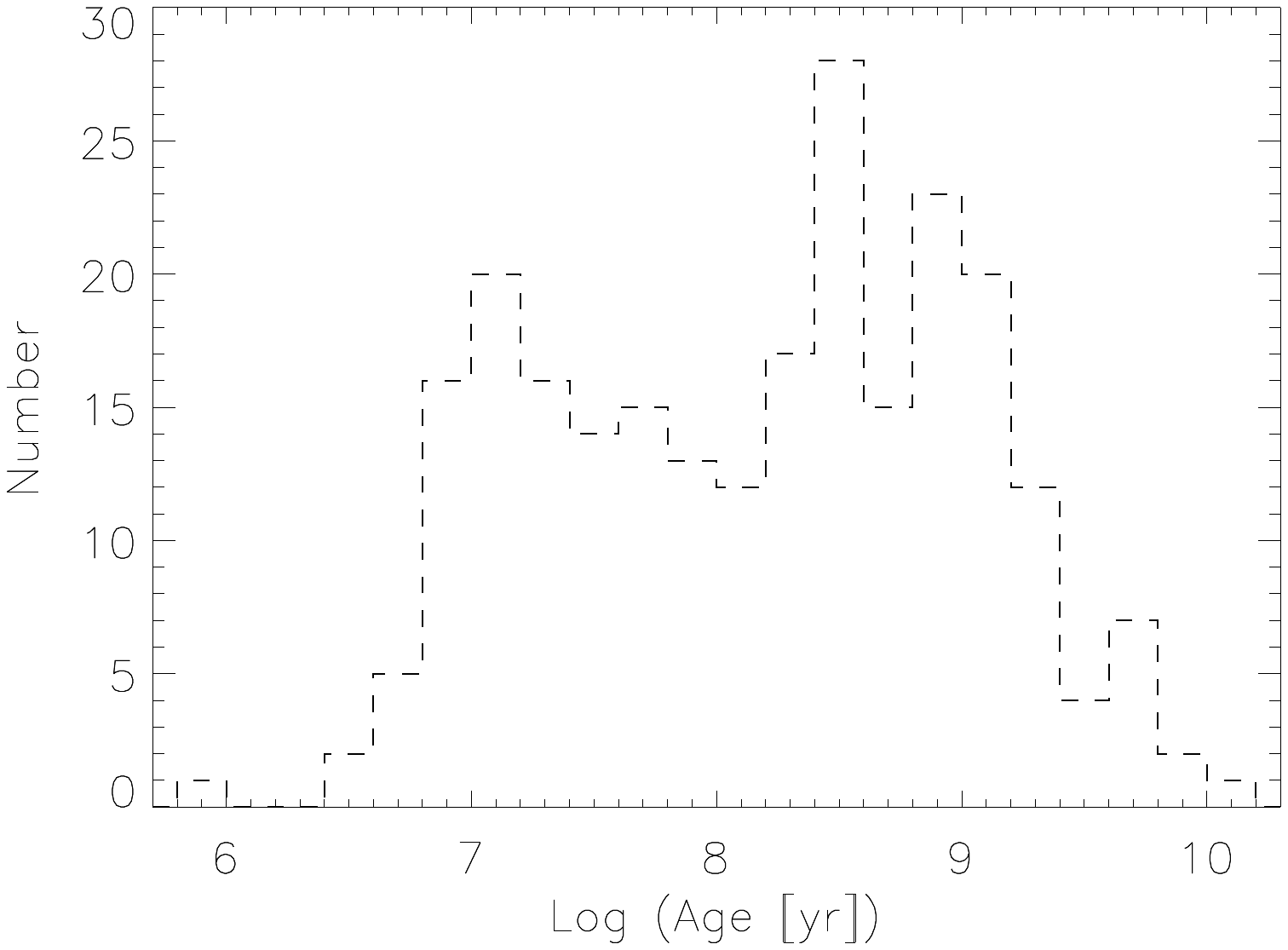}
\caption{\label{age_distribution} \textit{Left:} Histogram of the age
distributions of CCS\,1 -- old FSR clusters. \textit{Right:} Histogram of the
age distributions of CCS\, 2 -- WEBDA counterparts of FSR clusters.}
\end{figure*}

\subsection{Calibration of Cluster Model Distances}\label{distcal}

The model distances determined above are not necessarily accurate without
further calibration. \citet{Foster2012} have shown that distances estimated from
foreground star counts to dark clouds with maser sources agree with maser
parallax measurements in most cases. \citet{2012MNRAS.425.1380I} have used a
similar approach to us to measure distances to dark clouds with jets and
outflows. They used objects of known distances from the Red MSX Source (RMS, MSX
- Midcourse Space Experiment) survey by \citet{2008A&A...487..253U} to calibrate
their distances, which resulted in an accuracy for the distances that resembled
the intrinsic scatter of the calibration objects. 

For the FSR star cluster (candidates) the situation is more complex. Typically
the objects are not associated with dark clouds, hence the separation of
foreground and background objects (via median NIR colours) is less certain.
Furthermore, crowding in the cluster (and partly even in the control field) will
influence the observed foreground star density. Finally, it is unclear which
parameter values used to determine the model distance $d_{mod}$, lead to the
most accurate results.

Thus, we use sets of calibration samples of star clusters, a number of
corrections to the observed foreground star density and a parameter study to
investigate the accuracy of our method. In the following section we detail our
calibration approach.

\subsubsection{Cluster Calibration Samples}\label{ccs}

As in \citet{2012MNRAS.425.1380I}, we aim to have a Cluster Calibration Sample
(CCS) that consists of objects of a similar nature to the targets whose
distance we are aiming to determine. Hence, ideally we would use a sub-set of
the FSR clusters which have accurately determined distances. 

There are three obvious choices: i) CCS\,1 -- The sample of old open FSR clusters investigated
by \citet{2010MNRAS.409.1281F}; ii) CCS\,2 -- known FSR clusters that have a
counterpart in the WEBDA\footnote{\tt http://www.univie.ac.at/webda/} database
by \citet{1995ASSL..203..127M}; iii) CCS\,3 -- FSR cluster counterparts with
distances in the online version\footnote{\tt
http://www.astro.iag.usp.br/$\sim$wilton/} of the list of star clusters from
\citet{2002A&A...389..871D}.

All three of these samples have their obvious disadvantages. Both CCS\,2 and CCS\,3
are constructed from literature searches. The cluster distances in these samples
are determined by various methods and are hence inhomogeneous. Thus, it is
impossible to estimate the intrinsic scatter of the distances of the sample
clusters. CCS\,1, on the other hand, has homogeneously determined distances
(with a 30\,\% scatter, \citet{2010MNRAS.409.1281F}) but consists almost
exclusively of old star clusters. The full FSR sample, however, should 
contain a mix of young, intermediate age and old objects and it is unknown if
there are systematic differences if our method is applied to clusters of
different ages. 

Thus, we use two of the cluster samples to test and calibrate our distance
calculation method: i) CCS\,1 -- the old FSR\,clusters from
\citet{2010MNRAS.409.1281F}; ii) CCS\,2 -- the FSR counterparts in WEBDA, since
this catalogue generally includes only high accuracy measurements compared to the
list from \citet{2002A&A...389..871D}, which is more complete.

i) The old FSR sample contains 206 old open clusters. We remove every object that
is a suspected globular cluster (e.g. FSR\,0190 \citet{2008MNRAS.383L..45F} and
FSR\,1716 \citet{2008MNRAS.390.1598F}), which has less than 30 stars within one
core radius, and/or less than 30 stars in the control field (the latter two
conditions are for the A-sample of stars). We also remove any cluster with a core
radius of more than 0.05$^\circ$. This selection leaves 115 old open clusters in
the CCS\,1 sample. 

ii) We cross-match the entire FSR list with all entries in WEBDA. We consider the
objects as a match, if there was exactly one counterpart within 7.5\arcmin. In
the case where two FSR objects are near the same WEBDA entry the objects are
removed. We also require that every WEBDA match has a distance, age and reddening
value. Finally, as for CCS\,1, we remove objects with less than 30 stars within
one core radius, less than 30 stars in the control field (for the A-sample of
stars) and a core radius of more than 0.05$^\circ$. The final CCS\,2 sample after
these selections contains 241 clusters. 

The two CCSs have different properties which we will briefly discuss here. The
majority of clusters in CCS\,1 are between 1 and 4\,kpc distance, but there are a
number of objects that are up to 8\,kpc or further away. In contrast, CCS\,2
contains mostly clusters which are less than 3\,kpc distant, with very few
objects more than 5\,kpc away. The reddening distribution of CCS\,1 is biased
towards low A$_V$ clusters. The number of objects declines steeply with
increasing extinction, and there are almost no clusters with A$_V >$\,4\,mag. The
CCS\,2, on the other hand, has a homogeneous distribution of A$_V$ values between
0 and 3\,mag. Again, high extinction clusters are rare in the sample. However the
biggest difference between the two samples of calibration clusters is the age
distribution (see Fig.\,\ref{age_distribution}). While in CCS\,1 almost all
clusters are about 1\,Gyr old (within a factor of a few), CCS\,2 shows a more
homogeneous distribution of ages between 10\,Myrs and a few Gyrs. 

\subsubsection{Measuring Calibration Accuracy}\label{accuracy}

To quantify how well our method estimates the cluster distances for the
different CCSs we use the logarithmic distance ratio defined for each cluster as:

\begin{equation}
R = \log_{10}\left( \frac{d_{lit}}{d_{mod}} \right)
\end{equation}

Where $d_{lit}$ is the cluster distance as obtained from the literature and
$d_{mod}$ the distance determined from the foreground star density and the BGM.
If the value of $R$ is positive, then our our method underestimates the cluster
distance, if $R$ is negative, we overestimate the distance. We determine the
$rms$ of the distribution of $R$ values for each CCS and use this as a measure
for the scatter $S = (10^{rms} - 1)$ of our method. 

\subsubsection{Crowding and Extinction Correction}\label{correct_overcrowding}

There are a number of factors that we use to determine the model
distance to the cluster which will alter the measured foreground star
density (see Sect.\,\ref{id_foreground}). These are large scale foreground extinction and
crowding (in the general field and the centre of the cluster). Both effects are
not considered by the BGM, but can in principle be corrected for.

If there is large scale foreground extinction in the direction of the cluster,
the BGM will essentially predict a higher star density ($\rho_{BGM}$) than the
one we measure in the control field ($\rho_{con}$) around the cluster.
Similarly, close to the Galactic Plane and in particular near the Galactic
Centre, crowding becomes a major factor for the number of detected stars in
2MASS. Again, the star density predicted by the BGM will be larger compared to
the measured value in our control field. Thus, to correct for both effects
simultaneously, we multiply the measured foreground star density
$\rho^{cl}_{fg}$ by a factor:

\begin{equation}
X_1 = \frac{\rho_{BGM}}{\rho_{con}}
\end{equation}

Furthermore, crowding will be increased in the area of the cluster, since there
are naturally more stars in that region. The effect this has can be estimated,
since the observed star density in the control area should be equal to the
density of non-cluster stars in the cluster area. The latter can be easily
determined by summing up the non-membership probabilities of all stars in the
cluster area and dividing by $A_{cl}$. Thus, correcting the measured foreground
star density by another factor will correct for the additional crowding in the
cluster area compared to the control field: 

\begin{equation}
X_2 = \frac{\rho_{con} \times A_{cl}}{\sum\limits_i^{ } \left( 1.0 - P^i_{cl} \right)}
\end{equation}

This factor also addresses related differences in the completeness limit which is
used in the BGM simulation for the cluster, but has been estimated in the control
area. 

In summary,  to estimate the model distance for the clusters via the BGM we will
use the corrected foreground star density $\rho^{cl, cor}_{fg}$ which is
determined as:

\begin{equation}
\rho^{cl, cor}_{fg} = X_1 \times X_2 \times \rho^{cl}_{fg} = \frac{\rho_{BGM}
\times A_{cl}}{\sum\limits_i^{ } \left( 1.0 - P^i_{cl} \right)} \times \rho^{cl}_{fg}
\end{equation}

\begin{figure}
\includegraphics[width=8.6cm,angle=0]{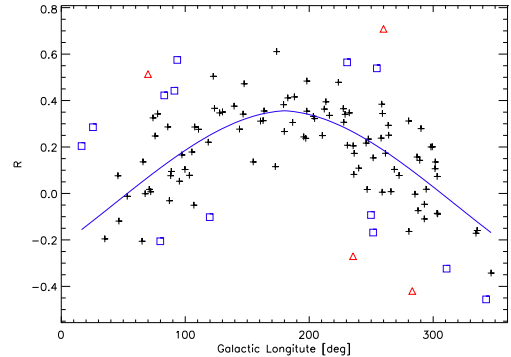}

\caption{ \label{R_l_2r25_sigma_example_plot} Plot of $R$ vs. the Galactic
longitude  of the CCS\,1 objects. These are based on distance calculations using
the A-sample, a cluster radius of  $1 \times r_{cor}$ and a nearest neighbour
number in the photometric decontamination of $N = 25$. The solid line indicates
our 3rd order polynomal fit. Black cross' represent points used for the fit with
$2 \sigma$ clipping, blue squares represent clusters that were excluded at the
$2 \sigma$ level and red triangles represent clusters excluded at the $3 \sigma$
level.}

\end{figure}

\begin{figure*}
\includegraphics[width=8.6cm,angle=0]{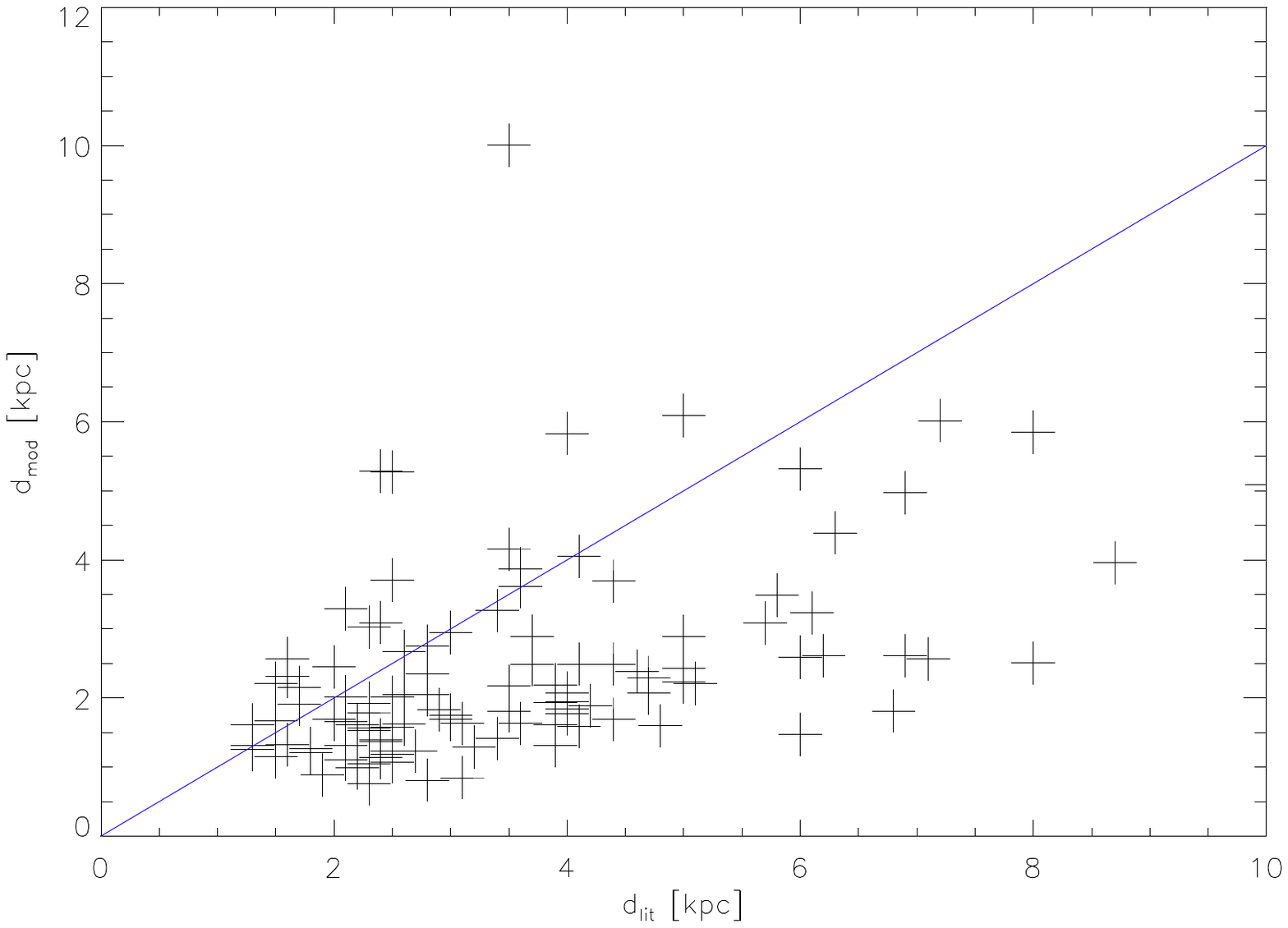} \hfill
\includegraphics[width=8.6cm,angle=0]{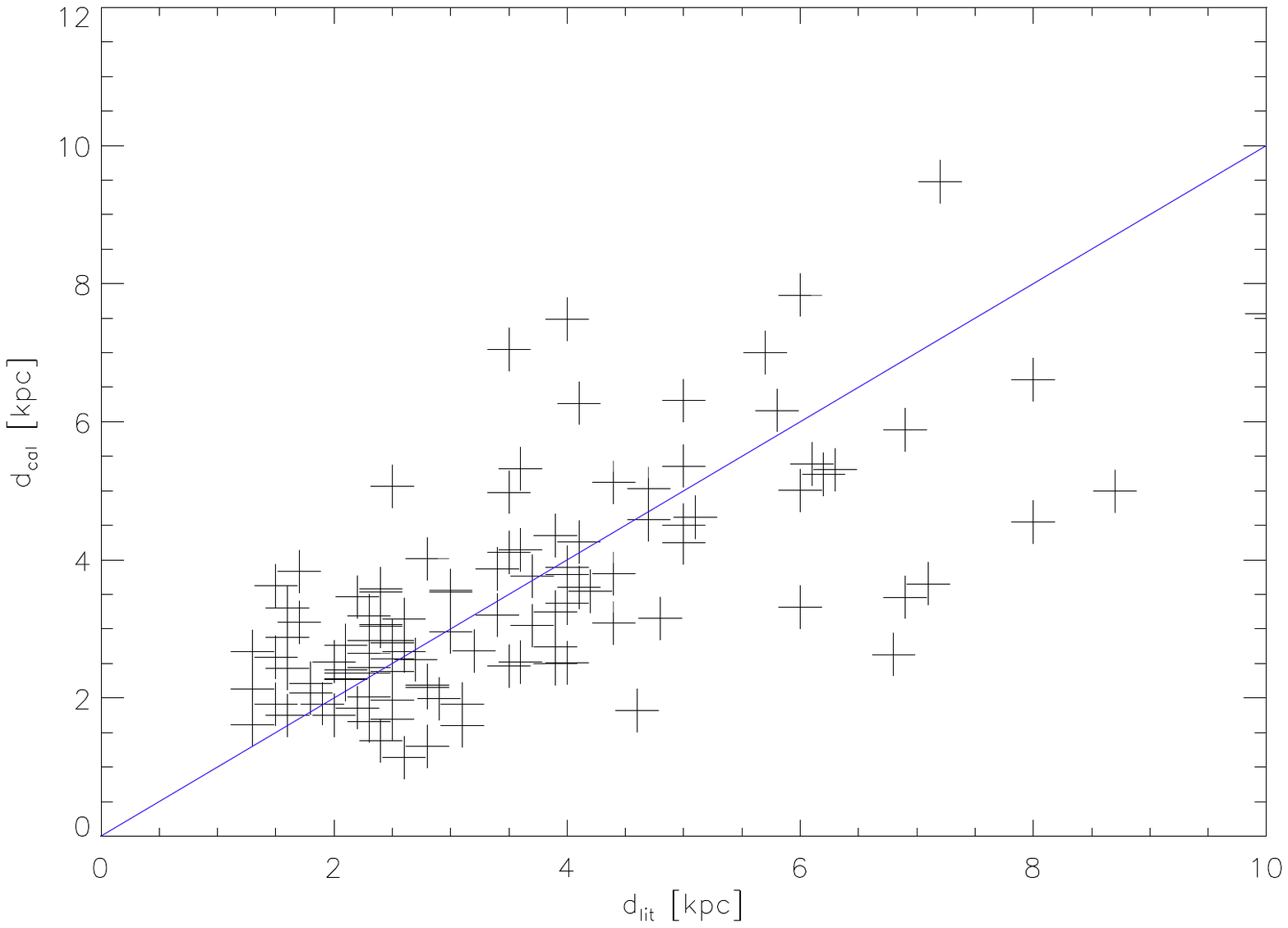} \hfill
\caption{\label{dcal_dlit_plot} Example plot of the literature distances
$d_{lit}$ for CCS\,1 clusters (black crosses) against our uncalibrated model
distance $d_{mod}$ (left) and the calibrated model distances $d_{cal}$ (right).
The calibration used the A-sample, a nearest neighbour in the photometric
decontamination of $N=25$ and a radius of $1 \times r_{cor}$. We overplot the
1:1 line in both cases. The final scatter $S_{cal}$ corresponds to a 40\,\%
uncertainty in the calibrated distances.}
\end{figure*}

\subsubsection{Position Dependent Corrections}\label{cali_calculations}

With the observed foreground star density corrected for crowding and large scale
extinction, we can estimate the model distance, $d_{mod}$, and compare to the
literature distance, $d_{lit}$, of the CCSs to investigate the scatter $S$ of the
distribution of the logarithmic distance ratios for all clusters in each CCS.

We investigate if there are any obvious correlations of the value of $R$ with
cluster parameters that are known or can be estimated for every cluster
(candidate) in the FSR sample. These include the Galactic Position ($l$ and
$b$), the apparent radius, the control field star density and the extinction
(based on $HK_{med}$, see Sect.\,\ref{id_foreground}). 

A significant trend occurs with the Galactic longitude. This can be seen in 
Fig.\,\ref{R_l_2r25_sigma_example_plot}. There a clear underestimate of the
cluster distances towards the Galactic Anticentre is apparent, while closer to the
Galactic Centre our method overestimates the distances. We find that a simple
3$^{rd}$ order polynomial fit can be used to correct this trend and hence to
decrease the apparent scatter $S$. Thus, we fit:

 %
 %

\begin{equation}\label{eq_calR}
R = C_{1} + C_{2} \times L + C_{3} \times L^{2} + C_{4} \times L^{3}
\end{equation}
 
with $L$\,=\,$\left| l - 180^\circ \right|$ and $l$ the Galactic longitude
of the cluster, and determine the parameters $C_i$ from the fit. We can then
calibrate our model distances by re-arranging Eq.\,\ref{eq_calR} and obtain the
calibrated distance:

 %
 %
 
\begin{equation}\label{eq_Distance}
d_{cal} = d_{model} \times 10^{ C_{1} + C_{2} \times L + C_{3} \times L^{2} +
C_{4} \times L^{3} }
\end{equation}
 
Note that we use a $3 \sigma$ clipping for the polynomial regression procedure
to remove obvious outliers. Once the correlation of $R$ with $l$ is corrected
for, none of the other above mentioned cluster parameters shows any dependence on
$R_{cal}$,  re-determined as 

\begin{equation}\label{eq_Distance_cal}
R_{cal} = \log_{10}\left( \frac{d_{lit}}{d_{cal}} \right). 
\end{equation}

We then determine the scatter $S_{cal}$ from the $rms$ of the $R_{cal}$
values to quantify the accuracy of our calibration procedure.  In
Fig.\,\ref{dcal_dlit_plot} we show an example of the improvements the calibration
makes to the distance estimates.

\begin{figure}
\includegraphics[width=8.6cm,angle=0]{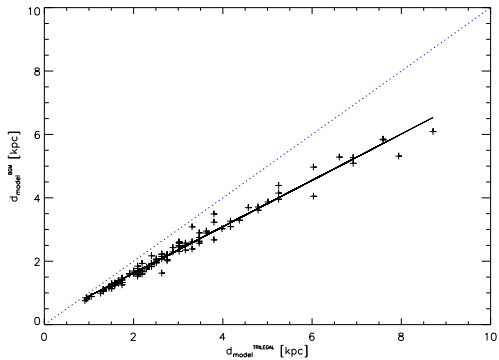} \hfill

\caption{\label{bgm_trilegal} Comparison of model distances $d_{model}$ obtained
from the Besan\c{c}on Galaxy Model and the TRILEGAL model. The crosses indicate
our calibration clusters from CCS\,1. All distances are estimated for the
A-sample of stars, a nearest neighbour number in the photometric decontamination
of $N = 25$ and a radius of $1 \times r_{cor}$. We overplot a blue dotted
one-to-one line as well as a black solid linear fit to the data.}

\end{figure}
 
We investigate if this calibration indeed removes the apparent dependence of our
calculated distances on the specific galactic model used. For this purpose we
compared the distances obtained from the Besan\c{c}on Galaxy Model
\citep{2003A&A...409..523R} with values estimated by using the
TRILEGAL\footnote{\tt http://stev.oapd.inaf.it/cgi-bin/trilegal} model from
\citep{2012rgps.book..165G, 2005A&A...436..895G}. Distances from the latter
model are obtained in the same way as described in Sect.\,\ref{distance_det} for
the BGM. In Fig.\,\ref{bgm_trilegal} we show one example of a comparison of the
two model distances $d_{model}$ obtained for the CCS\,1 sample of clusters. The
two distances depend linearly on each other, i.e. the TRILEGAL model predicts
distances which are a factor of about 1.3 times larger than the distance from
the BGM. However, since this is a linear relationship, our calibration described
above will entirely remove this difference. If we would use the TRILEGAL model
our calibration would simply result in a slightly different value for the
parameter $C_{1}$ in Eq.\,\ref{eq_calR}. This shows that our calibration
procedure removes any systematic dependence on the specific galactic model used,
since every other model will behave in a similar way. Thus, our distances are
model independent. We further note that the scatter in Fig.\,\ref{bgm_trilegal}
is only 5.5\,\%, i.e. much smaller than the final uncertainty of our distances
(see Sect.\,\ref{model_parameters}). Hence the contribution of the galactic
model to the statistical noise of our method is negligable.

\subsection{Optimisation of Distance Determination}\label{model_parameters}

The above described corrections and calibrations are performed for each
combination of the 'free' model parameters. Hence, we repeat the calibration
procedure for the A-sample and C-sample of stars, the two sets of CCSs, cluster
areas from one to three core radii and the nearest neighbour number (during the
membership probability calculations) from 10 to 30. 

For each set of free parameters we determine the scatter $S_{cal}$ to identify
the parameter set that leads to the best calibration, i.e. the lowest scatter.
Here we briefly discuss the results and implications of this optimisation
exercise.

\begin{itemize}

\item {\bf Photometric Samples:} For CSS\,1 there is a noticeable overall lower
value for the scatter $S_{cal}$ for the A-sample compared to the C-sample,  the
actual  size of which  depends on the other parameters.  However, for CSS\,2
there is no significant change in the scatter $S_{cal}$ between the A-sample and
the C-sample,  independent of other parameter values. As the aim of our project
is to find a method of distance determination for a homogeneous data set of
clusters, we choose the A-sample for our calibration and the final distance
calculations.

\item {\bf Photometric Decontamination:} The nearest neighbour number $N$ used
in the determination of the membership probabilities generates no trends or
significant differences in the scatter $S_{cal}$ for CSS\,1 and CSS\,2. In other
words, if we only vary $N$ and keep all other parameters the same, the scatter
$S_{cal}$ randomly fluctuates by a few percent, much less than the variations
caused by changes in one of the other parameters. This shows that the
decontamination procedure is independent of this parameter value (in the
investigated range of $10 \le N \le 30$) for our calibration samples. This is
understandable since $N$ only determines the resolution with which we can
identify potential cluster members in CCM space (as discussed in
Sect.\,\ref{cluster_prob}) and has no influence on the colours of the most
likely members which is used to determine the distances. To avoid any random
fluctuations due to the choice of $N$ (even if small) we hence choose two values
for this parameter ($N=15$ and $N=25$) for the distance calculations.

\item {\bf Cluster Area:} We vary the radius of the cluster area between one
and three times $r_{cor}$. As for the choice of $N$, we find that there is no
significant variation of the scatter $S_{cal}$ with cluster area. However, due
to the nature of the star density profile (Sect.\,\ref{find_radii}), the
majority of cluster members will be within 1$\times r_{cor}$, most cluster
members will be within 2$\times r_{cor}$, and the additional stars between 2
and 3$\times r_{cor}$ will be most likely field stars. Therefore, we only use
1$\times r_{cor}$ and 2$\times r_{cor}$ in the calibration and final distance
calculation.

\item {\bf Cluster Calibration Samples:} We spilt CCS\,2 into 'old'
(Log(Age[yrs]) $\geq 8.5$) and 'young' (Log(Age[yrs]) $< 8.5$) subsets,
CCS\,2$_{o}$ and CCS\,2$_{y}$ respectively. We then use these to investigate
if/how the scatter $S_{cal}$ varies with the mean age of the calibration sample.
We find that there is no statistically significant change in $S_{cal}$ produced
by CCS\,2 and CSS\,2$_{o}$. However, the scatter produced by CSS\,2$_{y}$
randomly fluctuates depending on the other parameters. Additionally, there is an
overall increase in the scatter when using CCS\,2$_{y}$ compared to the entire
CCS\,2 sample. This could be due to the fact that a younger cluster sample, such
as CCS\,2$_{y}$, will contain clusters which have a high fraction of YSOs with
high K-band excess. These stars could result in an inaccurate determination of
$HK_{med}$ (see below in Sect.\,\ref{id_yso}), and thus an erroneous number of
foreground stars which in turn can lead to a higher scatter in the calibration.
Therefore, our calibration should be applied to either a sample with a range of
ages or an older sample.  We choose CCS\,1, as it gives consistently a smaller
scatter $S_{cal}$ compared to CCS\,2. Based on the above discussion all
determined distances for clusters with a large fraction of YSOs, should be
treated with care.

\end{itemize}

In summary, we find that the A-sample of stars in conjunction with CCS\,1 leads
consistently to the smallest scatter for the distance calibration. We find no
significant or systematic influence of the radius of the cluster area or the
nearest neighbour in the photometric decontamination on the quality of our
method. 

We therefore choose four different sets of parameter values to determine the
distance to all FSR candidate clusters. The parameter values and the
corresponding scatter $S_{cal}$ for CCS\,1 and CCS\,2 are listed in
Table\,\ref{dis_calibrations}. The final distance for each cluster (listed in
Table\,\ref{app_table} in the Appendix) is then determined as the median of the
four distances from these calibration sets utilising CCS\,1. The typical scatter
and hence the accuracy of our distance calculation is better than 40\,\%,
considering the intrinsic scatter of 30\,\% for the distances of the calibration
sample \citet{2010MNRAS.409.1281F}.

\begin{table}

\caption{\label{dis_calibrations} Scatter $S_{cal}$\,[\%] of the four selected
calibration sets for CCS\,1 and CCS\,2. All calculations are done with the
A-sample of stars. The final cluster distances are determined as the median of
the four distances determined using the calibration for CCS\,1. {\it N} lists
the nearest neighbour number in the photometric decontamination and {\it Radius}
the distance out to which stars are included into the cluster area.}

\begin{center}
\begin{tabular}{lcccc}
\hline
 & {\it  N} & {\it Radius} & $S_{cal}$\,(CCS\,1) & $S_{cal}$\,(CCS\,2) \\
 &    & [$r_{cor}$] & [\%] & [\%] \\
\hline
\textbf{Set 1} &  15 & 1 & 36 & 54 \\  
\textbf{Set 2} &  25 & 1 & 37 & 50 \\   
\textbf{Set 3} &  15 & 2 & 40 & 55 \\
\textbf{Set 4} &  25 & 2 & 46 & 56 \\
\hline
\end{tabular}
\end{center}
\end{table}

\subsection{Identification of Young Stellar Objects}\label{id_yso}

We aim to identify and determine the fraction of YSOs ($Y_{frac}$) in each
cluster candidate. This is vital for the determination of the extinction values
(see Sect.\,\ref{ext_cal}), since disk excess stars can potentially lead to an
overestimate of the colour excess and thus the wrong extinction. Furthermore,
clusters with a significant fraction of YSOs are young. Thus, $Y_{frac}$ is
also an important age indicator for the cluster.  For each cluster candidate,
the $Y_{frac}$ calculation (as described here) is repeated for every
calibration set as described in Sect.\,\ref{model_parameters}.  

We identify YSOs in each cluster by calculating the reddening free parameter $Q$
for each star in the cluster area. We define $Q$ as:

\begin{equation} 
Q = (J-H) - \frac{E(J-H)}{E(H-K)} \times (H-K)
\end{equation}

Where $(J-H)$ and $(H-K)$ are the NIR colours of the star. We use a value of
$E(J-H) / E(H-K) = 1.55$ \citep{1990ARA&A..28...37M} for all determinations of
$Q$. We consider an object a disk-excess source (or YSO) if its $Q$ value is
smaller than $-$\,0.05\,mag by more than 1\,$\sigma$ (estimated from the
photometric uncertainties).  

Similar to the $HK_{med}$ determination (Sect.\,\ref{id_foreground}), we
calculate $Y_{frac}$ for the top 25\,\% to 45\,\% of the highest probability
member stars, in 5\,\% increments i.e. five $Y_{frac}$ values are determined for
each cluster and calibration set. The final value for $Y_{frac}$ listed in
Table\,\ref{app_table} in the Appendix is the median of the individual YSO
fractions averaged over the four calibration sets.

To determine the YSO fraction we count the potential YSOs in the cluster and the
total number of cluster members, $N_{YSO}$  and $N_{cl}$ respectively :

\begin{equation}
N_{YSO} = \sum\limits^{N}_{i, Q \le -0.05mag} P^i_{cl}
\end{equation}

\begin{equation}
N_{cl} = \sum\limits^{N}_{i} P^i_{cl}
\end{equation}

Where $N_{YSO}$ is the number of potential YSOs that are members of the cluster,
$N_{cl}$ is the number of all cluster members and $N$ is the number stars in the
top 25\,\% to 45\,\% of the highest probability cluster members.

The YSO fraction is in principle the ratio of $N_{YSO}$ and $N_{cl}$. However,
the number of YSOs is typically small and potentially influenced by photometric
scatter. We thus determine a limit for the YSO fraction by assuming that each
cluster contains at least $N_{YSO} - \sqrt{N_{YSO}}$ young stars:

\begin{equation}
Y_{frac} = \frac{N_{YSO} - \sqrt{N_{YSO}}}{N_{cl}}
\end{equation}

If $Y_{frac}$ is a negative value, i.e. there are less than one YSOs in the
cluster, the YSO fraction is set to zero.

\subsection{Extinction Calculation}\label{ext_cal}

In addition to distance, we aim to determine the extinction to all FSR cluster
candidates using the colour excess of the most likely cluster members we
identified in Sect.\,\ref{cluster_prob}. As for the determination of the
fraction of YSOs in the cluster (Sect.\,\ref{id_yso}), the extinction
calculation (as described here) is repeated for every calibration set as
described in Sect.\,\ref{model_parameters}.  The final extinction value for each
cluster candidate as quoted in Table\,\ref{app_table} in the Appendix is the
median of the four individual extinction values.

For our calculation we utilise two different colours:

i) ($H-K$), which we used in Sect.\,\ref{id_foreground} to identify foreground
stars; The ($H-K$) colour is less well suited to determine the extinction, since
its intrinsic value for stars depends on the spectral type as well as the
luminosity class of the object. Generally this colour can range from 0.0\,mag for
A-stars to 0.4\,mag for the latest spectral types (excluding rare OB stars and
L-type brown dwarfs). More typical average colours for our observed clusters are
between 0.1 and 0.3\,mag. The advantage of the ($H-K$) colour is that we can
determine it for every star in the cluster area.

ii) ($H-[4.5]$), which we estimate from the combination of WISE and 2MASS; The
intrinsic ($H-[4.5]$) colour is almost independent of spectral type and
luminosity class \citep{2011ApJ...739...25M} since it measures the slope of the
spectral energy distribution in the Rayleigh-Jeans part of the spectrum. Hence,
it has a well defined zero point. However, due to the lower resolution and depth
of WISE, we can determine this colour only for about half the stars in the
cluster area.

Since both utilised colours use the H-band, this is the natural pass band we
determine the extinction in. 

We calculate the H-band extinction
$A_{H}^{H45}$ from the ($H-[4.5]$) colour excess by means of:

\begin{equation}\label{ah45_equation}
A_{H}^{H45} = \frac{A_{H}}{A_{H}-A_{4.5}} \times \left< H - [4.5] \right>
\end{equation}

Where $\left< H - [4.5] \right>$ denotes the colour excess between the observed
$(H-[4.5])$ colour and the zero point. We utilise the zero point of
($H-[4.5]$)$_0 = 0.03$\,mag valid for main sequence stars
\citep{2011ApJ...739...25M} and the extinction law $A_{4.5}/A_{H} = 0.28$ from 
\citep{2005ApJ...619..931I}. 

We calculate the H-band extinction $A_{H}^{HK}$ from the ($H-K$) colour excess using:

\begin{equation}\label{ahk_equation}
A_{H}^{HK} = \frac{A_{H}}{A_{H}-A_{K}} \times \left< H - K \right>
\end{equation}

Where $\left< H - K \right>$ denotes the colour excess between the observed
($H-K$) colour (identical to $HK_{med}$, see Sect.\,\ref{id_foreground}) and the
zero point. We again utilise the extinction law from \citet{2005ApJ...619..931I}
which gives $A_{K} / A_{H} = 0.65$. To establish the zero point ($H-K$)$_{0}$ 
we plot $A_{H}^{H45}$ against $HK_{med}$ for all clusters with $A_{H}^{H45}
<$\,1\,mag and determined the off-set. We find that  ($H-K$)$_{0} = 0.06$\,mag.

This method will naturally overestimate the extinction for clusters with an
enlarged population of YSOs and hence disk excess emission stars. The excess
colour in these cases is caused by warm dust in the disk and not foreground
extinction. We identify clusters for which this may be an issue via the fraction
of intrinsically red sources in the NIR colour-colour diagrams (see
Sect.\,\ref{id_yso}). We consider all clusters with $Y_{frac}$ above 10\,\% as 
YSO clusters. These objects are excluded from the statistical analysis in this
paper, but note that the fraction of such clusters in the FSR sample is rather
low (see Sect\,\ref{young_results}).

\section{Results}\label{results}

Our calibration and optimisation procedure requires that for each cluster there
are at least 30 stars in the A-sample within one cluster core radius and that
the core radius is smaller than 0.05$^\circ$ (Sect\,\ref{cluster_prob}).  We
further exclude all known globular clusters and clusters where the distances
from the four calibration sets (see Sect.\,\ref{model_parameters}) scatter by
more than 3\,$\sigma$ compared to the average cluster. These requirements
result in 1017 FSR objects being excluded, of which 931 are open clusters or
candidates.

We determine the distance to 771 of the FSR objects i.e. 43\,\% of the entire
FSR sample, of which 377 are known open clusters and 394 are new cluster
candidates. Based on the total numbers of these objects in the FSR list (681
known clusters and 1021 new cluster candidates), these are 55\,\% and 39\,\%,
respectively. This reflects the fact that in comparison to the new FSR cluster
candidates, the known clusters should have, on average, more members and hence
will more likely pass our selection criteria. The new FSR cluster candidates are
typically less pronounced and a fraction of about 50\,\% of them might not be a
real cluster \citep{2007MNRAS.374..399F}. This has also been confirmed by the
analysis of spatially selected sub-samples of the FSR cluster candidates by
\citet{2008MNRAS.385..349B} and \citet{2010A&A...521A..42C}. They find that
about half the investigated candidates are star clusters with a tendency that
overdensities with less members are more likely not real clusters. We finally
note that the current version of the open cluster database\footnote{\tt
http://www.astro.iag.usp.br/$\sim$wilton/} from \citet{2002A&A...389..871D}
contains at the time of writing 141 of the 1022 FSR cluster candidates which
have been followed up by the community. Hence, even without a complete and
systematic analysis of the entire sample, 15\,\% of the clusters have already
been identified as real objects. All of this evidence indicates that at least
three quarters of all the confirmed clusters and cluster candidates investigated
here are real, or that the contamination of our cluster sample is less than
25\,\%.

Further to the distances, we calculated the extinction and YSO fraction for 775
cluster candidates (43\,\% of the FSR sample, excluding globular clusters). The
split of this group into known clusters and new candidates is 378 and 397
objects, respectively, or 56\,\% and 39\,\% of the entire cluster sample.  In
the following we will analyse the statistical properties of the entire sample.
We will not aim to gauge which of the individual cluster candidates is real or
not. This will be done in a forthcomming paper after the ages of the candidates
have been determined. However, to aid the reader in evaluating the
trustworthyness of individual properties we add two columns to
Table\,\ref{app_table} in the Appendix. We list as $N^{tot}_{2r_{core}}$ the
number of stars in the A-sample within two core radii for each cluster. We
furthermore sum up all membership probabilities ($P^i_{cl}$) of these stars to
determine the total estimated number of cluster members $N^{mem}_{2r_{core}}$ in
the same area. As a guide we note that the contrast, i.e. the ratio of cluster
members to field stars, is on average twice as high for the known clusters than
for the new candidates. This is a clear indication that more pronounced objects
are more likley to be real clusters.

\begin{figure*}
\includegraphics[width=8.6cm,height=7.2cm,angle=0]{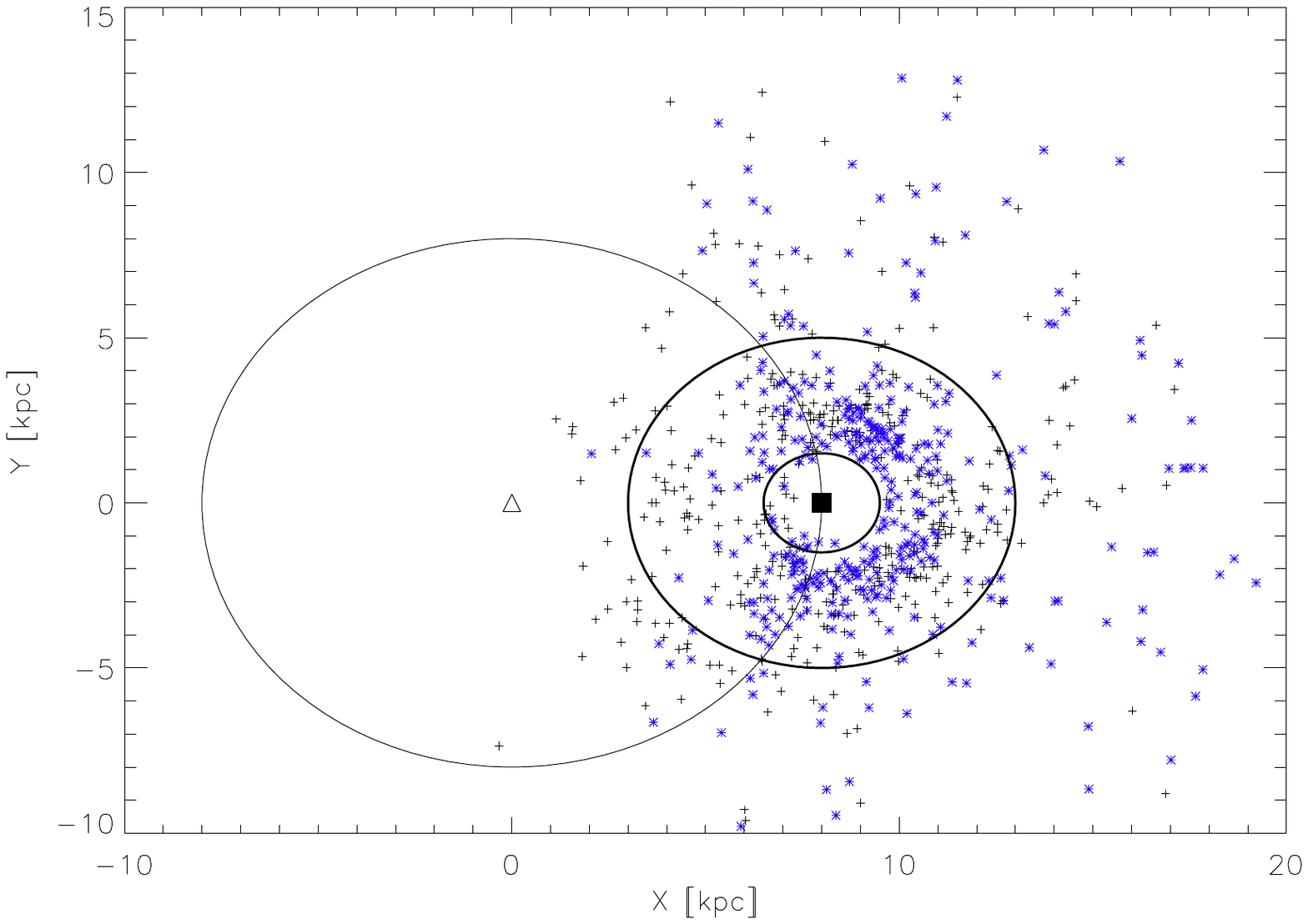} \hfill
\includegraphics[width=8.6cm,height=7.2cm,angle=0]{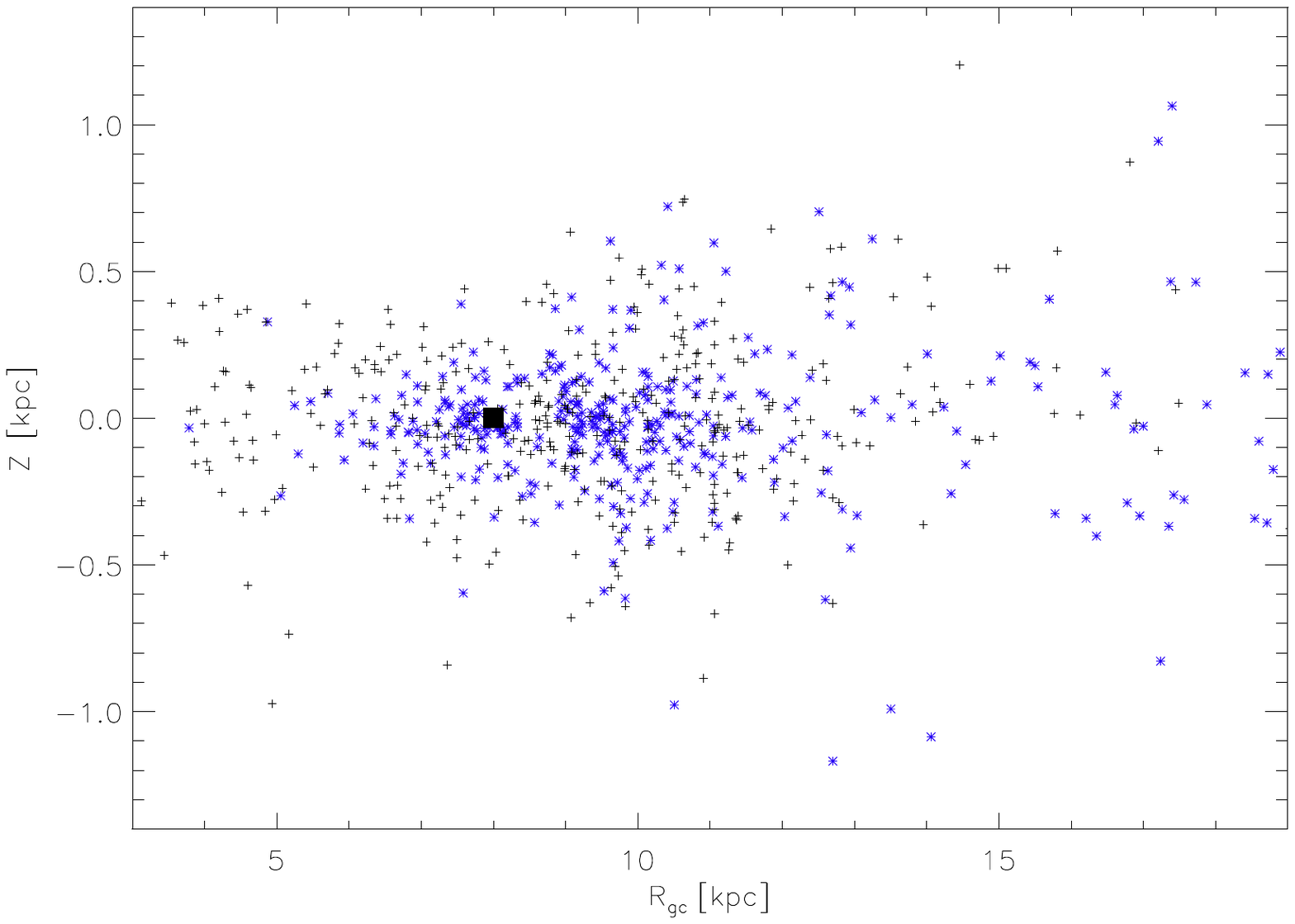}

\caption{\label{distance_distribution} \textit{Left:} Plot of the distribution
of FSR clusters in the Galactic Plane based on the distances determined in this
paper. The Galactic Centre is represented by a triangle. The smaller of the
three circles indicates a distance of 1.5\,kpc from the Sun, the medium sized
circle indicates a distance of 5\,kpc from the Sun, and large circle indicates
a distance of 8\,kpc from the Galactic Centre. \textit{Right:} Plot of the
height, Z, above and below the Galactic Plane of our clusters as a function of
the Galactocentric distance, $R_{gc}$. In both plots black crosses represent
new cluster candidates, blue stars represent previously known open clusters and
the Sun is represented by a black square.}

\end{figure*}

\subsection{Cluster distribution in the Galactic Plane}\label{GP_distr}

We investigate how the clusters with distances are distributed in the Galactic
Plane (GP) and determine completeness limits and selection effects. In
Fig.\,\ref{distance_distribution} we show the positions of all clusters
projected into the GP and the position with respect to the GP as a function of
the Galactocentric distance. Note that for all these plots we assume a distance
of 8\,kpc of the Sun to the Galactic Centre.

The density of clusters per unit area in the GP for our sample peaks at about
3\,kpc distance from the Sun. We find that averaged over the entire longitude
range there are about 15 star clusters per square kiloparsec projected into the
GP at a distance of 3\,kpc from the Sun in our sample. At 4\,kpc distance the
density drops to about half this level. Also towards smaller distances, a
similar significant drop in the density of clusters is found. The latter is
caused by the well known selection effect in the FSR list, which includes only
clusters of a certain apparent radius \citet{2010MNRAS.409.1281F}. Thus, many
nearby clusters were not included in the original FSR sample, or have been
excluded by our condition that the cluster core radius should be smaller than
0.05$^\circ$ (see Sect.\,\ref{ccs}). The drop in numbers at larger distances is
in part caused by the same effect -- the clusters are too compact to be included
-- and more distant objects are also too faint. 

The scale height of the clusters with respect to the GP has been calculated for
various sub-samples of our clusters. We excluded all clusters that have a
distance of more than 5\,kpc, since we are more likely to detect clusters
further away from the mid-plane at large distances due to extinction. The known
clusters in our sample have a scale-height of 235\,$\pm$\,20\,pc while the new
cluster candidates have a distribution with a width of 315\,$\pm$\,30\,pc. We
also split our sample in potentially younger and older clusters via the
fraction of YSOs (old: $Y_{frac} <$\,1\,\%; young $Y_{frac} >$\,1\,\%). Then we
find for the young clusters a scale-height of 190\,$\pm$\,15\,pc, and for the
older clusters 300\,$\pm$\,20\,pc.  These scale heights are in between the
numbers for clusters older and younger than 1\,Gyr in the FSR sample (older:
375\,pc; younger 115\,pc; \citet{2010MNRAS.409.1281F}). It is also larger than
the scale height of the dust in the solar neighbourhood (125\,pc
\citep{Drimmel2003,Marshall2006}). The potentially younger clusters in our
sample show a higher degree of concentration to the GP. In Paper II we will
determine the ages of all clusters and candidates to investigate in more detail
the age dependence of the scale height.

As can be seen in the original FSR paper \citep{2007MNRAS.374..399F} and also
amongst the old FSR clusters \citep{2010MNRAS.409.1281F}, the distribution of the
clusters along the GP is not homogeneous. In particular towards the GC the sample
is incomplete. This is most likely caused by crowding and low contrast between
cluster and field stars. For the older clusters, some of this could be a real
effect since old open clusters seem to exist in lower numbers at Galactocentric
distances smaller than that of the Sun \citep{Friel1995}. 

Here we utilise Kolmogorov-Smirnov tests (see e.g. \citet{1983MNRAS.202..615P})
to investigate if subsamples of the FSR clusters have similar distributions or
if they are different. We compare five different samples:

\begin{itemize}

\item[S1] Clusters with $Y_{frac} < 5$\,\%; old clusters

\item[S2] Clusters with 5\,\%\,$< Y_{frac}$; young clusters

\item[S3] Known open clusters amongst the FSR sample

\item[S4] New FSR cluster candidates

\item[S5] All FSR clusters, except known globulars

\end{itemize}

We determine the probability that any two of the samples are randomly drawn from
the same parent distribution. We also test the clusters in each sample against a
homogeneous distribution (SH). For this we only use clusters that are more than
60$^\circ$ away from the GC, since we know that close to the GC the FSR list is
incomplete \citep{2007MNRAS.374..399F}. We summarise the results in
Table\,\ref{ks_results}. 

These tests do reveal that most of the samples have a different distribution
along the GP. In particular, the sample of potentially old clusters (S1) seems to
be different from most of the other sub-samples. Note that probabilities larger
than 1\,\% and less than 99\,\% in the KS-test are not significant.

\begin{table}
\begin{center}

\caption{\label{ks_results} KS-test probabilities that two of our cluster
sub-samples are drawn from the same parent distribution. S1: older clusters,
$Y_{frac} < 5$\,\%; S2: young clusters, 5\,\%\,$< Y_{frac}$; S3: Known FSR
clusters; S4: New cluster candidates; S5: All FSR clusters, except known
globular clusters; SH: A homogeneous distribution along the GP -- but more than
60$^\circ$ away from the GC.}

\begin{tabular}{lcccccc}
\hline
 &  S1 & S2 &S3 & S4 & S5 & SH \\
\hline
S1 &   --- & 0.25 & 0.02 & 2.53 & 0.43 & 33.8 \\  
S2 & 0.25 &   --- & 39.5 & 4.07 & 9.89 & 17.2 \\  
S3 & 0.02 & 39.5 &  --- & 0.51 & 15.4 & 52.2 \\   
S4 & 2.53 & 4.07 & 0.51 &  --- & 44.6 & 69.6 \\
S5 & 0.43 & 9.89 & 15.4 & 44.6 &  --- & 76.0 \\
SH & 33.8 & 17.2 & 52.2 & 69.6 & 76.0  &  --- \\
\hline
\end{tabular}
\end{center}
\end{table}

\begin{figure}
\includegraphics[width=8.6cm,angle=0]{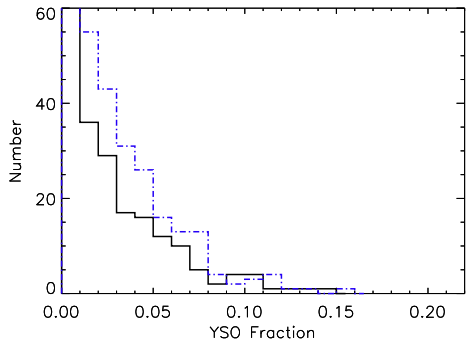}

\caption{\label{histogram_yso} Histogram showing the distribution of $Y_{frac}$
of our sample. The solid black line represents new cluster candidates and the
dot-dash blue line the known open clusters. The peak in the first bin ($Y_{frac}
< $\,1\,\%) is too high to be shown -- 258 new cluster candidates and 165 known
open clusters.}

\end{figure}

\subsection{Young Clusters}\label{young_results}

In Paper\,II we will determine the ages of all our cluster candidates and
investigate the age distribution as well as how the cluster properties such as
scale height change with age (as discussed above in Sect.\,\ref{GP_distr}). So
far the only measure of age we have is the fraction of young stars ($Y_{frac}$)
in each cluster as determined in Sect.\,\ref{id_yso}. In
Fig.\,\ref{histogram_yso} we show the distribution of the YSO fraction for all
clusters. 

There is a steep decrease of the number of clusters with increasing YSO fraction
and only a small number of cluster candidates have a high $Y_{frac}$. The sample
is dominated by clusters with essentially zero YSOs. Only 18 cluster candidates
have $Y_{frac} > $\,10\,\%, and no objects have $Y_{frac} > $\,20\,\%. The
apparent lack of clusters with $Y_{frac} >$\,20\,\% in our sample means there
are potentially no clusters with an age of less than 4\,Myrs -- according the
relation of YSO fraction and age for embedded clusters from \citet{Lada2003}.
There are two obvious reasons for this: i) the FSR list is assembled from NIR
2MASS overdensities, hence will naturally lack young and embedded objects. ii)
We do not include L-band excess stars in our YSO fraction, and our YSO fraction
is a lower limit. Using the L-band would require us to include the WISE
photometry, and we have seen in Sect.\,\ref{ext_cal} that only about half the
2MASS sources in each cluster can be reliably cross-matched to their WISE
counterparts.

The four clusters with the highest YSO fraction are FSR\,0488 (Teutsch\,168),
FSR\,1127 (NGC\,2311), FSR\,1336 (a new cluster candidate) and FSR\,1497
(Ruprecht\,77). 

\begin{figure*}
\includegraphics[width=8.6cm,angle=0]{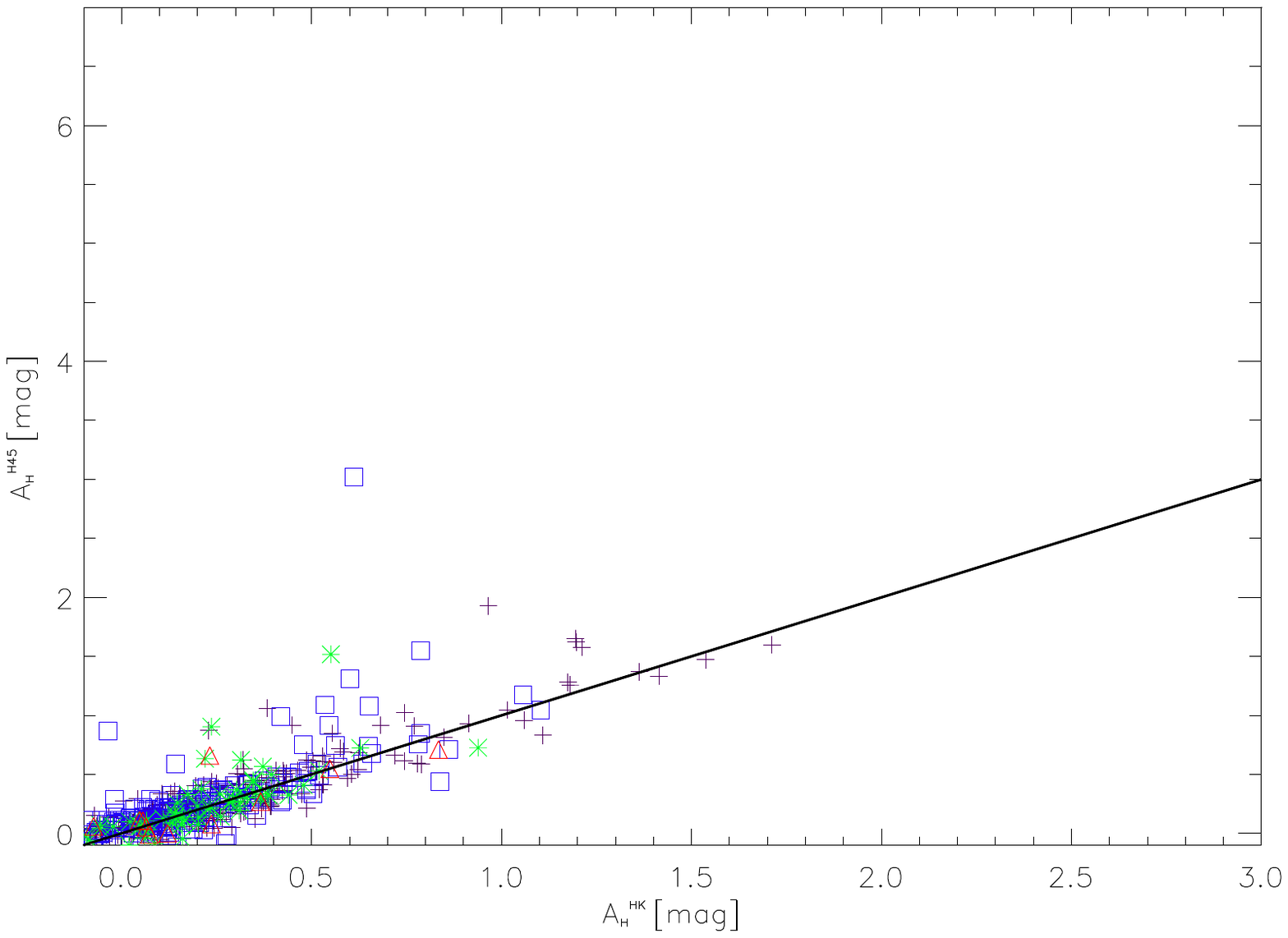} \hfill
\includegraphics[width=8.6cm,angle=0]{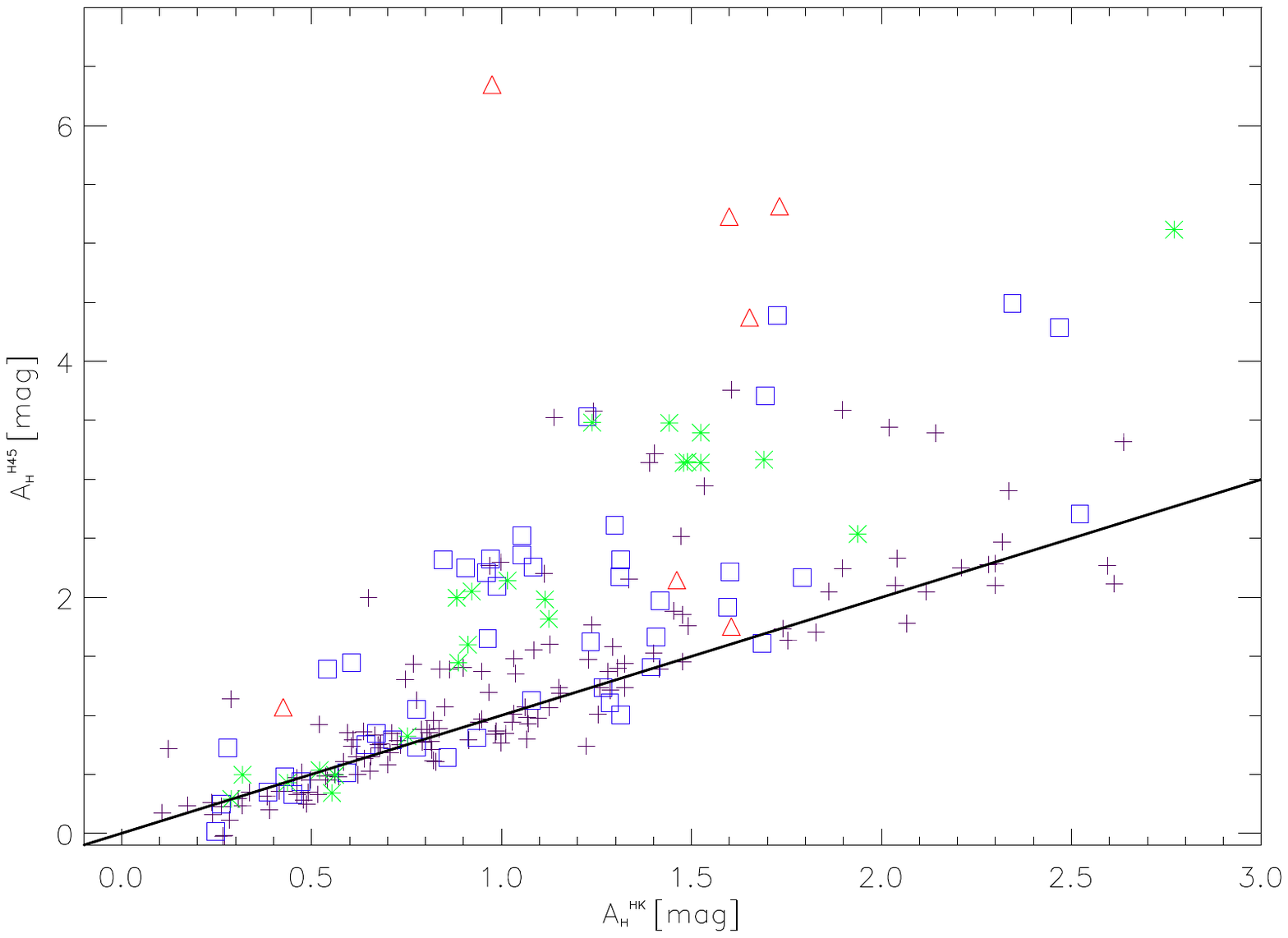}

\caption{\label{figure_ahk_ah45} Plot of $A_{H}^{HK}$ against $A_{H}^{H45}$.
Red triangles represent clusters with $Y_{frac}>0.10$, green stars
are clusters with $0.05 < Y_{frac} < 0.10$, blue squares are clusters with $0.01
< Y_{frac} < 0.05$ and purple crosses are clusters with $Y_{frac}<0.01$. The left
panel includes all clusters with distances $d_{cal} < 5$\,kpc and the right panel
all clusters more distant than 5\,kpc or where we did not estimate a distance but
the extinction.}

\end{figure*}

\subsection{Extinction Law}

As outlined in Sect.\,\ref{ext_cal}, we determine a zero point of
$(H-K)_0$\,=\,0.06\,mag by comparing the cluster extinction values determined
from $(H-K)$ and $(H-[4.5])$ colour excess and excluding clusters with
$Y_{frac}>0.10$.  In Fig.\,\ref{figure_ahk_ah45} we show how the H-band
extinction values calculated from the two colour excesses compare to each other.
The vast majority of clusters follow the one-to-one line with a $rms$ of
0.18\,mag in $A_H$. This even holds for clusters where we were unable to
determine a reliable distance. 

There are a number of clusters for which the extinction calculated from the
$(H-[4.5])$ excess is significantly higher than that calculated from the $(H-K)$
excess. These differences are most likely due to the fact that only the brightest
$[4.5]$\,$\mu$m sources are matched to 2MASS and hence the extinction calculation
is biased towards intrinsically redder objects. We exclude all clusters more than
3\,$\sigma$ above the one-to-one trend line  from our subsequent analysis of the
dependence of the extinction on the cluster distance.

We estimate how the extinction per unit distance changes for our cluster sample
as a function of Galactic longitude. To establish a reliable value, we apply the
following restrictions to our cluster sample: i) All clusters have to be closer
than 5\,kpc to the Sun in order to ensure a reliable distance; ii) The fraction
of YSOs should be smaller than 10\,\% for the extinction not to be influenced by
disk excess emission stars; iii) Clusters have to be closer than 150\,pc to the
GP to ensure the line of sight is mostly in the Galactic midplane; iv) The
extinction of the cluster determined from $(H-[4.5])$ and $(H-K)$ excess differ
by less than 3\,$\sigma$ compared to the one-to-one trend line.

\begin{figure*}
\includegraphics[width=8.6cm,angle=0]{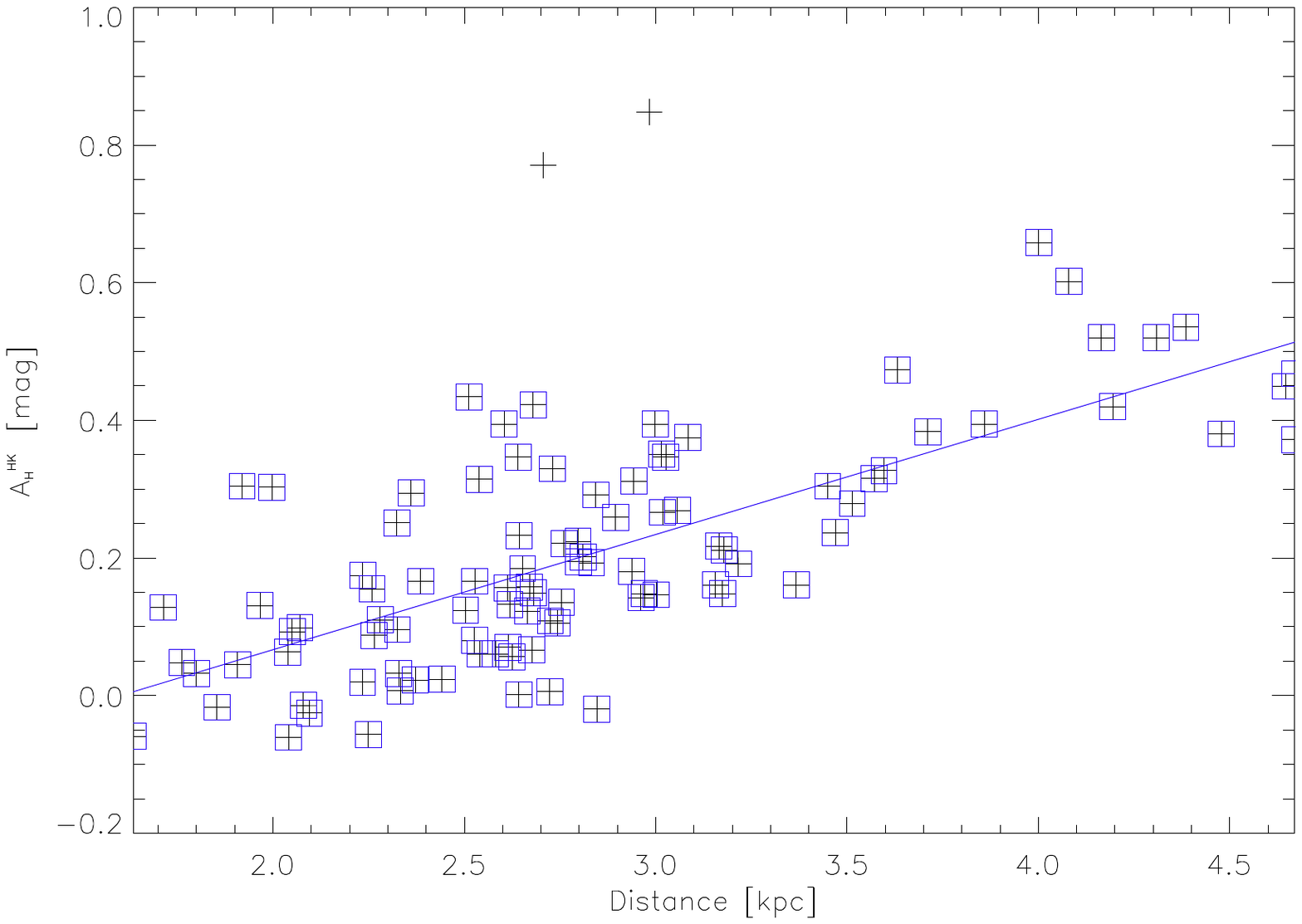} \hfill
\includegraphics[width=8.6cm,angle=0]{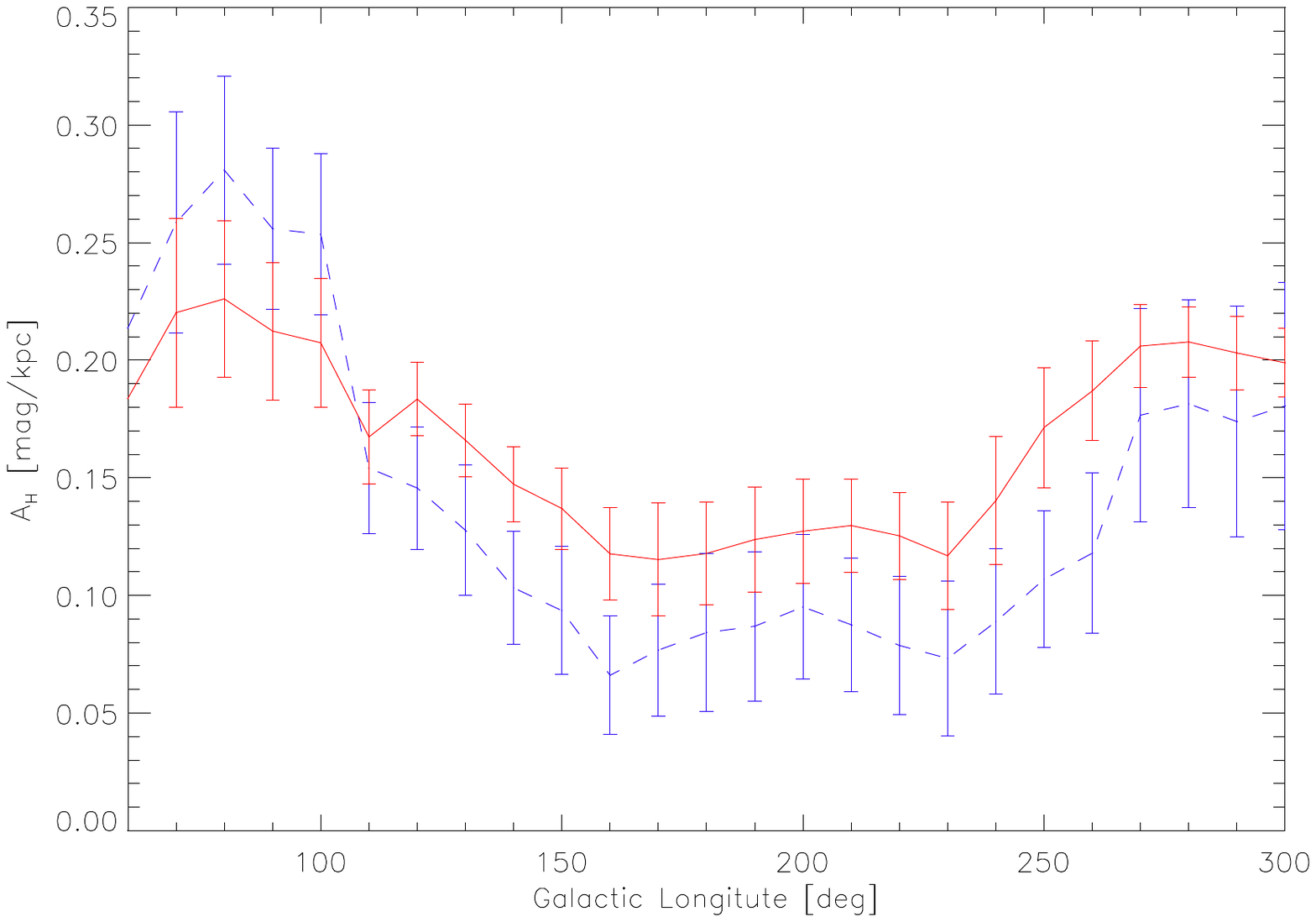}

\caption{\label{figure_ah_distance}\textit{Left:} The plot shows $A_{H}^{HK}$
against our calibrated distance values for clusters with
$l=110^{\circ}\pm30^{\circ}$. Crosses represent clusters in this region, and
boxed crosses are clusters which were not excluded from the fit by our $3\sigma$
clipping. \textit{Right:} This shows the H-band extinction per unit distance as a
function of Galactic longitude. The red solid line represents $A_{H}^{HK}$ per
kpc and the blue dashed line $A_{H}^{H45}$ per kpc.}

\end{figure*}

In the left panel of Fig.\,\ref{figure_ah_distance} we show an example of how
$A_H^{HK}$ depends on the distance for our clusters. Only clusters at $l = 110
\pm 30^\circ$ are shown. The overplotted solid line is the best linear fit,
boxed crosses indicate the clusters used, whilst un-boxed crosses mark the
clusters excluded by a 3\,$\sigma$ clipping. The fit indicates an extinction law
of 0.17\,mag/kpc extinction in the H-band. Converted to optical extinction using
\citet{1990ARA&A..28...37M} this is $A_V$\,=\,0.9\,mag/kpc. 

We investigate how this extinction law changes as a function of Galactic
longitude in the right panel of Fig.\,\ref{figure_ah_distance}. We determine
the extinction per unit distance every 10$^\circ$ of Galactic longitude, but
include all clusters that are within 30$^\circ$. We find that there is a
systematic dependence of the extinction per unit distance. The lowest values
($A_H =$\,0.1\,mag/kpc) are found near the Galactic Anticentre and a steady
increase is observed towards the GC where almost $A_H =$\,0.3\,mag/kpc is
reached. We only determined the extinction law if there are at least 10
clusters available in the longitude bin. It is important to note, that in
regions closer than 60$^\circ$ to the GC, very few clusters are available.
Figure\,\ref{figure_ah_distance} also shows that the general behaviour of the
extinction values is similar for both $A_H^{HK}$ and $A_H^{H45}$; they are
mostly indistinguishable within the uncertainties.

The canonical value of the extinction per unit distance, $A_V =$\,0.7\,mag/kpc
(e.g. \citet{2010MNRAS.409.1281F}), converts to $A_H =$\,0.12\,mag/kpc. This is
in agreement with our estimates for large regions towards the Galactic
Anticentre, but does not hold for lines of sight closer towards the GC. There,
values at least two or three times the canonical value seem to be more
appropriate. We find that the longitude dependence of the extinction can best
be described as:

\begin{equation}
A_H(l) {\rm [mag/kpc]} = 0.1 + 0.001 \times \left| l - 180^\circ \right| 
/^\circ ,
\end{equation}

for lines of sight of more than $60^\circ$ from the Galactic Centre. Our
calibration with respect to the Galactic longitude (see
Sect.\,\ref{cali_calculations}) is thus only needed because we used a constant
$A_V$/kpc in the BGM. If one employs our position dependent value for the ISM
extinction in the model, then only the crowding and extinction corrections
(Sect.\,\ref{correct_overcrowding}) are required. In principle this could
ensure the usability of our method without a calibration sample.

\subsection{Potential Improvements}\label{distance_method_improv}

One consequence of using 2MASS data to identify stars foreground to the clusters
is that the data has a resolution of about $2"$ \citep{2006AJ....131.1163S}.
Thus, we are unable to resolve close (apparent) binaries and to detect faint
cluster members. In other words, crowding in dense cluster centres and generally
near the GC or the Galactic Plane is the limiting factor in our accuracy.

With the availability of deeper homogeneous photometry data sets, such as
UKIDSS-GPS \citep{Lucas2008} and VISTA-VVV \citep{Minniti2010} better NIR data
will be available for studies of large samples of clusters (e.g.
\citet{2012A&A...545A..54C}).  These deeper and higher resolution data will not
only allow the detection of fainter cluster members, it also allows a more
accurate determination of the density of stars foreground to the cluster, since
deeper observations generally detect more foreground dwarfs than background
giants. Furthermore, these data will also allow us to study more compact and
distant  clusters, which are currently excluded due to the low spatial
resolution.

\section{Conclusions}

We present an automatic calibration and optimisation method to estimate distances
and extinction to star clusters from NIR photometry alone, without the use of
isochrone fitting. We employed the star cluster decontamination procedure from
\citet{2007MNRAS.377.1301B} and \citet{2010MNRAS.409.1281F} to calculate the
number of stars foreground to clusters and the Besan\c{c}on Galaxy Model
\citep{2003A&A...409..523R} to estimate uncalibrated model distances. 

Our method has been calibrated using two calibration sets of known open
clusters. We utilise a homogeneous list of 206 old ($>$100 Myr) open clusters
from the FSR list \citep{2010MNRAS.409.1281F} as well as all WEBDA entries
amongst the FSR clusters. Due to the nature of the entries in the WEBDA database
the latter calibration set is inhomogeneous, but covers a larger range of ages.
We find that the older FSR clusters provide the best calibration sample and we
achieve a distance uncertainty of less than 40\,\% after the calibration. Our
calibration procedure ensures that all our results are independent of the
specific galactic model used.

Using our calibration method, we determined the distances to 771 cluster
candidates from the FSR list. Of these, 377 were known open clusters and 394 are
new cluster candidates. We also determine the extinction to 775 clusters,  378
known and 397 new candidates. Based on the Q-parameter we also estimate the YSO
fraction for each cluster in our sample.

The spatial density of clusters per unit area in the GP for our sample peaks at
about 3\,kpc distance from the Sun. Thus, our sample is biased towards clusters
of this distance and lacks more distant and closer objects. This is caused by the
well known selection effect in the FSR list, which includes only clusters of a
similar apparent radius \citep{2010MNRAS.409.1281F}.  The scale height of the
younger clusters (YSO fraction above 1\,\%) is 190\,pc, while the remaining
(older) clusters show a scale height of 300\,pc.

We investigated how the extinction per unit distance to the clusters changes as
a function of Galactic longitude. There is a systematic dependence that can at
best be described by $A_H(l)$\,[mag/kpc]\,=\,0.1\,+\,0.001$\times \left| l -
180^\circ \right| /^\circ$ as long as the clusters are more than 60$^\circ$ from
the GC and not embedded in Giant molecular clouds. We suggest use of this
extinction law, instead of a constant canonical value for any simulations
performed with the Besan\c{c}on Galaxy Model. In turn this allows to use our
procedure for distance estimates without the need for a calibration sample. In
particular, for clusters in the GP but not projected onto Giant Molecular
Clouds, the $\left< H - [4.5] \right>$ colour excess can be used as a distance
indicator -- at least statistically.

\section*{Acknowledgements}
 
The authors would like to thank L.\,Girardi and B.\,Debray for their support in
accessing the galactic models via a script rather than the web interface.
A.S.M.\,Buckner acknowledges a Science and Technology Facilities Council
studentship and a University of Kent scholarship. This publication makes use of
data products from the Two Micron All Sky Survey, which is a joint project of
the University of Massachusetts and the Infrared Processing and Analysis
Center/California Institute of Technology, funded by the National Aeronautics
and Space Administration and the National Science Foundation. This research has
made use of the WEBDA database, operated at the Institute for Astronomy of the
University of Vienna. This publication makes use of data products from the
Wide-field Infrared Survey Explorer, which is a joint project of the University
of California, Los Angeles, and the Jet Propulsion Laboratory/California
Institute of Technology, funded by the National Aeronautics and Space
Administration.

\bibliographystyle{mn2e}
\bibliography{references}

\label{lastpage}

 %
 %

\onecolumn

\begin{appendix}



\include{appendix_table2}

\end{appendix}

\end{document}

%% file: appendix_table2.tex
\section{FSR Cluster Property Table}\label{appendix1}

\begin{longtable}{|l|c|rr|rr|r|rr|r|rr|}

\caption{\label{app_table} Summary table of the FSR cluster properties
determined with our method. The table lists the FSR ID number, the cluster type
(known open cluster or new cluster candidate), the Galactic coordinates and RA
and DEC (J2000), the distance in kiloparsec, the $H$-band extinction values
calculated from $H-K$ excess ($A_H^{HK}$) and from $H - [4.5]$ excess
($A_H^{H45}$),  the YSO fraction ($Y_{frac}$), the number of stars within two
core radii of the cluster centre ($N^{tot}_{2r_{core}}$), and the total number
of members (calculated as the sum of the membership probabilities of all stars)
in the same area ($N^{mem}_{2r_{core}}$). }

\label{fsr_table}    \\

\hline
\multicolumn{1}{|c|}{\textbf{FSR}} &
\multicolumn{1}{c|}{\textbf{Type}} &
\multicolumn{1}{c|}{\textbf{l}} &
\multicolumn{1}{c|}{\textbf{b}} &
\multicolumn{1}{c|}{\textbf{RA}} &
\multicolumn{1}{c|}{\textbf{DEC}} &
\multicolumn{1}{c|}{\textbf{d}} &
\multicolumn{1}{c|}{\textbf{$A_{H}^{H45}$}} &
\multicolumn{1}{c|}{\textbf{$A_{H}^{HK}$}}  &    
\multicolumn{1}{c|}{\textbf{$Y_{frac}$}} &
\multicolumn{1}{c|}{\textbf{$N^{tot}_{2r_{core}}$}} &
\multicolumn{1}{c|}{\textbf{$N^{mem}_{2r_{core}}$}}\\
\multicolumn{1}{|c|}{\textbf{ID}} &
\multicolumn{1}{|c|}{\textbf{}} &
\multicolumn{1}{c|}{\textbf{[deg]}} &
\multicolumn{1}{c|}{\textbf{[deg]}} &
\multicolumn{1}{c|}{\textbf{(J2000)}} &
\multicolumn{1}{c|}{\textbf{(J2000)}} &
\multicolumn{1}{c|}{\textbf{[kpc]}} &
\multicolumn{1}{c|}{\textbf{[mag]}} &
\multicolumn{1}{c|}{\textbf{[mag]}}   & 
\multicolumn{1}{|c|}{\textbf{ }}  &
\multicolumn{1}{|c|}{\textbf{ }}  &
\multicolumn{1}{|c|}{\textbf{ }}  \\
\hline  \endfirsthead

\multicolumn{10}{c}%

 {{\bfseries \tablename\ \thetable{} -- continued from previous page}}     \\ \hline
\multicolumn{1}{|c|}{\textbf{FSR}} &
\multicolumn{1}{c|}{\textbf{Type}} &
\multicolumn{1}{c|}{\textbf{l}} &
\multicolumn{1}{c|}{\textbf{b}} &
\multicolumn{1}{c|}{\textbf{RA}} &
\multicolumn{1}{c|}{\textbf{DEC}} &
\multicolumn{1}{c|}{\textbf{d}} &
\multicolumn{1}{c|}{\textbf{$A_{H}^{H45}$}} &
\multicolumn{1}{c|}{\textbf{$A_{H}^{HK}$}}  &  
\multicolumn{1}{c|}{\textbf{$Y_{frac}$}} &
\multicolumn{1}{c|}{\textbf{$N^{tot}_{2r_{core}}$}} &
\multicolumn{1}{c|}{\textbf{$N^{mem}_{2r_{core}}$}}\\
\multicolumn{1}{|c|}{\textbf{ID}} &
\multicolumn{1}{|c|}{\textbf{}} &
\multicolumn{1}{c|}{\textbf{[deg]}} &
\multicolumn{1}{c|}{\textbf{[deg]}} &
\multicolumn{1}{c|}{\textbf{(J2000)}} &
\multicolumn{1}{c|}{\textbf{(J2000)}} &
\multicolumn{1}{c|}{\textbf{[kpc]}} &
\multicolumn{1}{c|}{\textbf{[mag]}} &
\multicolumn{1}{c|}{\textbf{[mag]}}   &  
\multicolumn{1}{|c|}{\textbf{ }}  &
\multicolumn{1}{|c|}{\textbf{ }}  &
\multicolumn{1}{|c|}{\textbf{ }} \\
\hline  \endhead

\hline \multicolumn{12}{|r|}{{Continued on next page}}     \\ \hline
\endfoot

\hline
\endlastfoot

0001 &  new  & 0.03   &  03.47 &  17:32:22.2 &  -27:03:41  &  4.4 &   0.91  &  1.06 &   0.00 & 1015& 342 \\
0002 &  new  & 0.05   &  03.44 &  17:32:32.4 &  -27:03:52  &  4.3 &   0.84  &  1.05 &   0.00 &  365& 111 \\
0009 &  new  & 1.86   & -09.52 &  18:28:29.9 &  -31:53:49  &  3.4 &   0.05  &  0.32 &   0.00 &  286&  73 \\
0018 &  new  & 5.34   &  05.41 &  17:37:41.4 &  -21:33:27  &  4.1 &   0.41  &  0.53 &   0.00 & 1333& 337 \\
0019 &  new  & 5.52   &  06.08 &  17:35:39.3 &  -21:02:50  &  3.8 &   0.27  &  0.44 &   0.00 &  654& 233 \\
0022 &  new  & 6.18   &  00.84 &  17:56:28.8 &  -23:11:36  &  6.3 &   1.01  &  0.99 &   0.00 &  456& 237 \\
0023 &  new  & 6.58   &  00.78 &  17:57:35.1 &  -22:52:34  &   -  &   0.81  &  1.28 &   0.00 &  139&  34 \\
0025 &  new  & 7.54   &  05.65 &  17:41:43.6 &  -19:34:14  &  3.6 &   0.40  &  0.52 &   0.00 &  419& 118 \\
\end{longtable}